\documentclass[1p,review,sort&compress]{elsarticle}
\usepackage{amssymb,amsmath}
\usepackage{bm,enumerate,tikz,flowchart,txfonts}
\usepackage[colorlinks,citecolor=magenta,linkcolor=blue]{hyperref}
\usetikzlibrary{arrows}
\journal{Computer Physics Communications}

\newcommand{\rstate}[1]{\vert #1 \rangle}

\newcommand{\avstate}[3]{\langle #1 \vert #2 \vert #3 \rangle }
\newcommand{\dotstate}[2]{\langle #1 \vert #2  \rangle }

\begin{document}

\begin{frontmatter}

\title{Natural Orbital-Based Lanczos Method for Anderson Impurity Models}

\author[ruc,keylab]{Sheng Bi\corref{cor1}}
\cortext[cor1]{Email address: rucbsplu@ruc.edu.cn.}
\address[ruc]{Department of Physics, Renmin University of China, 100872 Beijing, China}
\address[keylab]{Beijing Key Laboratory of Opto-electronic Functional Materials and Micro-nano Devices (Renmin University of China) }

\author[stspclab]{Li Huang}
\address[stspclab]{Science and Technology on Surface Physics and Chemistry Laboratory, P.O. Box 9-35, Jiangyou 621908, China}

\author[ruc,keylab]{Ning-Hua Tong\corref{cor2}}
\cortext[cor2]{Email address: nhtong@ruc.edu.cn.}

\begin{abstract}
We implement the Lanczos algorithm on natural orbital basis to solve the zero-temperature Green's function of Anderson impurity models, following the work of Y. Lu, M. H\"{o}ppner, O. Gunnarsson, and M. W. Haverkort, Phys. Rev. B {\bf 90} (2014) 085102. We present the technical details, generalize the algorithm to the cases of particle-hole asymmetry, with local magnetic field, and of two impurities. The results are benchmarked with conventional Lanczos, quantum Monte Carlo, and numerical renormalization group methods, demonstrating its potential as a powerful impurity solver for the dynamical mean-field theory.
\end{abstract}


\end{frontmatter}

\section{Introduction}	

The Anderson impurity model (AIM)~\cite{Anderson} is one of the basic models in condensed matter physics. It describes the physics of a local electron orbital with on-site Coulomb repulsion embedded in a conduction electron band and is widely used to describe the dilute magnetic impurities in metals~\cite{Hewson}, Kondo effect~\cite{Kondo}, as well as impurity quantum phase transitions~\cite{Vojta}. In the past two decades, stimulated by the development and application of the dynamical mean-field theory (DMFT)~\cite{Vollhardt,Georges}, the study of AIM receives revived attention because in DMFT, a lattice Hamiltonian for the correlated electrons is mapped into an AIM with self-consistently determined electron bath. The core calculation of DMFT is the iterative solution of the self-energy of an AIM with arbitrary hybridization function. The AIMs generalized from single impurity or one bath to multiple impurities and/or multiple baths are also the target of active researches, both for describing physical impurity systems~\cite{Nozieres,Surer} and for solving the cluster extensions of DMFT equations~\cite{Maier}.

Given the importance of AIM and the lack of rigorous solution for general situations, it is naturally desirable to have an accurate, fast, and flexible method for solving the AIM, which is a challenging quantum many-body problem. There have been a variety of numerical approaches to solve AIM, each with its advantages and disadvantages. The exact diagonalization (ED)~\cite{Noack} and Lanczos~\cite{Dagotto} methods produce the exact self-energy of AIM with a finite number of discrete bath sites. Due to the exponential increase of Hilbert space dimension with system size, these methods are limited to small number of bath and impurity orbitals. Ideas such as the distributional ED~\cite{Granath} have been explored to overcome this problem. Quantum Monte Carlo methods, including the Hirsch-Fye~\cite{Hirsch} and various continuous time (CT-QMC) algorithms~\cite{Gull}, are essentially exact and flexible but face difficulties at very low temperatures and at calculating Green's functions (GFs) on real frequencies. The numerical renormalization group (NRG) method~\cite{Wilson} has extremely high accuracy at low energies but lacks resolution at high energies and is limited to small number of bath bands~\cite{Pruschke1,Mitchell} or impurities~\cite{Jones}. The recently developed hierarchical equation of motion method~\cite{Li} is highly efficient and versatile, but the required computing resources increase fast with decreasing temperature and with increasing number of the Lorentzians used to decompose the hybridization function. Analytical methods are also investigated, such as perturbation theories~\cite{Yamada,Dai,Tong}, non-crossing approximation~\cite{Bickers} and its extensions~\cite{Pruschke2}, equation of motion of GFs~\cite{Lacroix}, all with partial success. 

Recently, a series of studies disclosed an interesting feature of the ground state of AIM. That is, the ground state of AIM can be efficiently described by a limited number of Slater determinants formed on the optimal one-electron basis, the natural orbital (NO) basis~\cite{Lowdin,Bender}. This feature was employed to design highly efficient numerical algorithms for calculating the ground state and zero-temperature GFs of AIM. The configuration-interaction (CI) solver of AIM based on adaptive basis was explored in Ref.~\cite{Lin}. The natural orbital renormalization group algorithm~\cite{He1,He2} was developed to iteratively refine the NOs in a way similar to the restricted active space approach in quantum chemistry~\cite{Lin}. $O(N_{b}^{3})$ scaling of the computing cost with the number of bath sites $N_{b}$ is obtained~\cite{He2} and the study of a $2 \times 2$ cluster with $N_{b}=60$ was reported. In another work, the Lanczos algorithm based on sparse storage of NOs is designed and integrated into the DMFT self-consistent calculations~\cite{Lu}. The results obtained using $N_{b}=301$ are compared with the results from NRG, demonstrating the superior advantage of this method compared to traditional ED or Lanczos methods. Recently, the variational determination of the optimal electron orbital was demonstrated on the one- and five-orbital AIMs~\cite{Schuler}. 

In this paper, we study the NO-based Lanczos method proposed in Ref.~\cite{Lu}. The purpose is first to provide algorithm details that are important for the implementation of the code but lacking in the original work. Second, we extend this method to the cases of particle-hole asymmetry, under local magnetic field, and of two impurities. In all the cases, we demonstrate the accuracy and applicability of this method. The rest part of this paper is organized as follows. In Section~\ref{Anderson impurity model} we introduce the model that we study. In Section~\ref{natural orbital basis} and \ref{Orbital transformation}, the natural orbital basis is defined for the impurity model. The algorithm details about NO-based Lanczos and its difference from the conventional Lanczos method are given in Section~\ref{Impurity solver}. The GF is calculated in section \ref{Green's Function}. Section~\ref{results} presents the results from NO-Lanczos and compares them with NRG and CT-QMC results, including the results for two-impurity AIM. A summary is given in Section~\ref{Conclusion}.

\section{Anderson impurity model\label{Anderson impurity model}}   
We consider a general $N_d$-impurity AIM with the following Hamiltonian
\begin{equation} 
\label{AIMH}
H = H_{cond} + H_{imp} + H_{hyb}.
\end{equation}
The first part $H_{cond}$ describes the non-interacting bath,
\begin{equation}
H_{cond} = \sum_{k=1}^{N_{b}}\sum_{\sigma} {\epsilon_k c_{k \sigma}^\dagger c_{k \sigma}}.
\end{equation}
The second term 
\begin{equation}
\label{impurity term}
H_{imp} = -\mu \sum_{i=1}^{N_d} \sum_{\sigma} {{n}_{i \sigma}} + U \sum_{i=1}^{N_d} n_{i \uparrow} n_{i \downarrow} + \sum_{i<j} U_{ij} n_{i} n_{j}
\end{equation} 
describes $N_d$ local impurities with on-site Coulomb repulsion $U$ and inter-impurity interaction $U_{ij}$. $n_i = \sum_{\sigma} n_{i\sigma}$ is the electron number operator of impurity site $i$. The impurities are coupled to bath electrons via the hybridization term
\begin{equation} 
\label{hybrid term}
 H_{hyb}=  \sum_{i=1}^{N_d} \sum_{k,\sigma} v_{ik} \left( d_{i \sigma}^{\dagger}c_{k \sigma} + d_{i\sigma}^{\dagger} c_{k \sigma} \right).
\end{equation}
The hybridization function matrix is defined as 
\begin{equation} 
\label{hybrid matrix}
   \Gamma_{ij}(\omega) \equiv \displaystyle\sum_{k} \frac{v_{ik} v_{kj}}{\omega - \epsilon_k}.
\end{equation}
In this paper, we first consider the diagonal hybridization function matrix
\begin{equation}   \label{hybrid Lorentzian}
 \Gamma_{ij}(i\omega_n)= \delta_{ij} \frac{\pi \Delta \omega_c}{i \omega_n + i \omega_c \text{sgn}(\omega_n)},
\end{equation}
which corresponds to the Lorentzian spectral function on the real frequency axis, $-1/\pi \text{Im} \Gamma_{ii}(\omega + i \eta) = \Delta \omega_c^2/ (\omega^2 + \omega_c^2)$. The hybridization matrix with both diagonal and off-diagonal elements will be studied in Fig.\ref{Fig14}. Throughout this paper, we use $\omega_c=1.0$ as the unit of energy. $\Delta$ is the hybridization strength.

\begin{figure}[t]
\centering
   \includegraphics[width=0.8\textwidth]{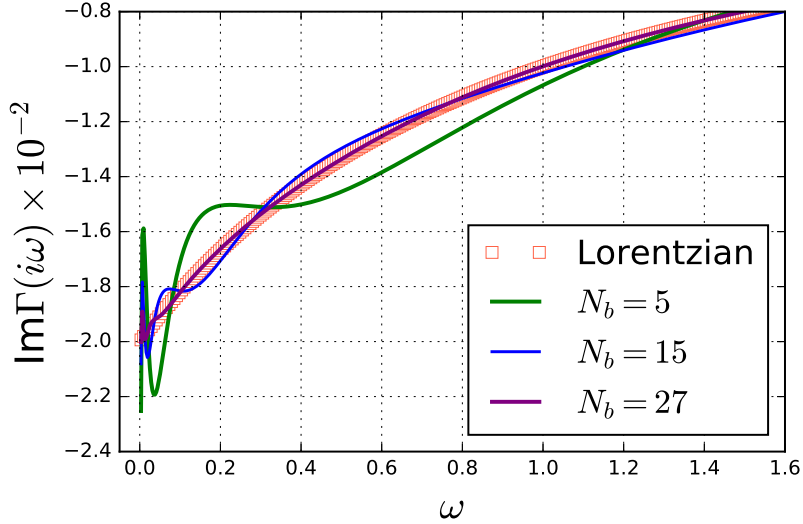}
\caption{(Color online). Fitting of Lorentzian hybridization function Eq.~(\ref{hybrid Lorentzian}) with $N_d=1$ (red square) using different number of bath sites $N_{b}=5$, $15$, and $27$ (lines). The real part of $\Gamma(i \omega_n)$ is zero. Here, $\omega_c = 1$, $\pi \Delta=0.02$, and $\beta=1000$ is used in the Matsubara frequencies. \label{Fig_fit} }
\end{figure}

The Hamiltonian parameters $v_{ik}$'s and $\epsilon_k$'s used in this paper are obtained by the least square fitting of Eq.~(\ref{hybrid Lorentzian}) using Eq.~(\ref{hybrid matrix}) on the Matsubara axis, as done in most ED impurity solvers~\cite{Caffarel}. A factor $1/\omega_n^{s}$ could be added to the cost function to enhance the fitting accuracy in the low frequency regime.
Considering that the relatively large number of bath sites used in this work already gives small fitting error, here we use the simplest fitting scheme without $1/\omega_n^{s}$ factor. In this paper, we will first consider the single impurity case $N_d=1$ and then extend our study to $N_d=2$. The effect of fitting is shown in Fig.~\ref{Fig_fit} for $N_d=1$ and  $N_{b}=5$, $15$, and $27$. The fitting is already very accurate for $N_{b} = 15$ and excellent for $N_{b}=27$.

\section{Nature orbital basis   \label{natural orbital basis}}  

A many-electron state $| \Psi \rangle$ can be expanded into the linear combination of Slater determinants defined by various occupancies of single particle orbitals $\{ | \phi_{i} \rangle = c_{i}^{\dagger} |0 \rangle \}$. Here $|0\rangle$ is the vacuum state and $c_{i}^{\dagger}$ creates an electron on the orbital $| \phi_{i} \rangle$. The average electron number $n_{i} = \langle \Psi | c_{i}^{\dagger} c_{i} | \Psi \rangle$ measures the probability of $\phi_{i }$ being occupied in $|\Psi \rangle$. Those orbitals with $n_i \sim 1$ have a large probability of being occupied in each Slater determinant, while those with $n_{i} \sim 0$ being probably empty. Therefore, the appearance and disappearance of such orbitals are fixed in the Slater determinants of $| \Psi \rangle$. Various occupancies of the partially occupied orbitals $0 < n_{i} < 1$ generate the active space, which contains the Slater determinants required for an accurate expansion of $|\Psi \rangle$. 

Among all the single particle orbitals, NO has the most extreme distribution of occupancies $n_i$ and hence allows for the smallest active space. By using the NO basis one needs the least number of Slater determinants to represent $\Psi$ to a given precision~\cite{Helgaker,He1}. NO $\psi_i$ is defined as the eigenstate of the one-particle density matrix $\bm{\gamma}_{\sigma}$, i.e., $\bm{\gamma}_{\sigma} \psi_{i \sigma} = n_{i \sigma} \psi_{i \sigma}$. Here, $\boldsymbol{\gamma}_{ij\sigma} = \langle \Psi |c_{i \sigma}^{\dagger} c_{j \sigma} | \Psi \rangle$ and $\{ c_{i \sigma}^{\dagger} \} $ and $\{c_{i \sigma}\}$ are the creation and annihilation operators of electrons on a set of orthonormal spin-orbitals $\{ \phi_{i\sigma} \}$. In this paper, we only use the single-particle density matrix that is diagonal in the spin index $\sigma$.

\begin{figure}[t]
\centering
   \includegraphics[width=0.8\textwidth]{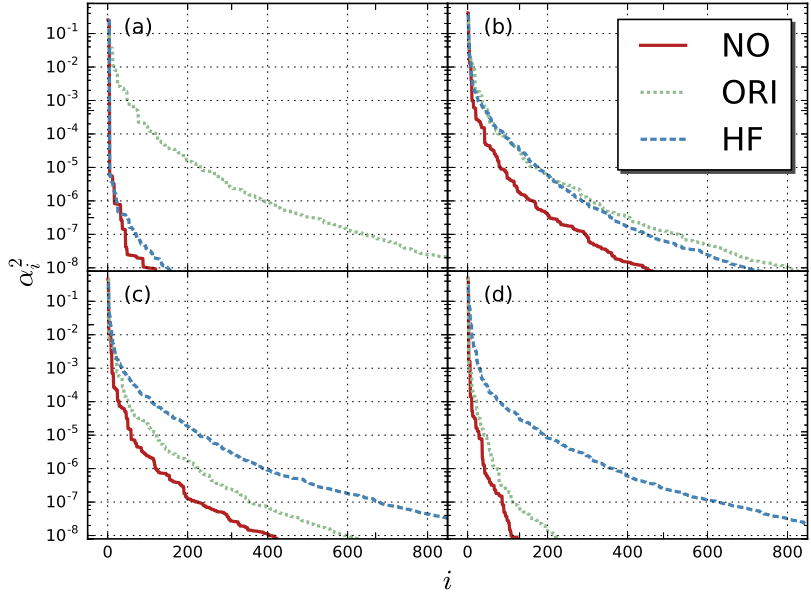}
\caption{(Color online). The probability $\alpha_i^2$ of the Slater determinant $|\phi_i \rangle$ in the ground state of AIM, sorted in descending order. The AIM contains one impurity site and $7$ bath sites.  ORI and HF represent the original and the Hartree-Fock basis, respectively. (a) $U=0.2 \pi \Delta$; (b) $U=10 \pi \Delta$; (c) $U= 20 \pi \Delta$; (d) $U=100 \pi \Delta$. Other parameters are $ \pi \Delta=0.02$, $\mu=U/2$. \label{Fig_slater}} 
\end{figure}

Previous works~\cite{Lin,He1,He2} show that the ground state of AIM can be efficiently represented on the basis of NO basis because the number of active orbitals is on the order of the number of interacting sites $N_d$, much smaller than the total number of orbitals $N_d + N_{b}$. The required number of Slater determinants is much smaller than the full dimension of the Fock space $4^{N_d + N_{b}}$. This makes it possible to significantly reduce the computational cost for Lanczos calculation of the ground state. In Fig.~\ref{Fig_slater}, we expand the ground state of AIM into a linear combination of Slater determinants and show the distribution of the probability (coefficient squared). It is done for the exact ground state of AIM with $N_d=1$ and $N_{b}=7$ for increasing $U$ values at particle-hole symmetric point $\mu=U/2$ (from (a) to (d)). Each panel contains results of three different single-particle bases: NO basis, original basis on which the Hamiltonian Eq.~(\ref{AIMH}) is defined, and Hartree-Fock (HF) basis from diagonalizing $\gamma_{\sigma}$ of the HF ground state. For all three bases, only a small fraction of the total $4^{8}=65536$ Slater determinants contribute significantly to the ground state ({\it i.e.,} with probability larger than $10^{-8}$). For all $U$ values, the NO basis always gives the steepest decaying curve and the ground state contains less than $400$ significant Slater determinants on the NO basis. For small $U$ values, the curve for the HF basis decays faster than that of the original basis, close to that of the NO basis. For large $U$ values, the curve for the HF basis decays slowest.

\section{Orbital transformation\label{Orbital transformation}} 
we diagonalize the single-particle density matrix $\bm{\gamma}_{\sigma}$, 
\begin{equation}
   \bm{U}_{\sigma}^{-1} \bm{\gamma}_{\sigma} \bf{U}_{\sigma} = \mathbf{\Lambda}_{\sigma}.
\end{equation}
Here, $\bm{U}_{\sigma}$ is an unitary matrix and $\left( \mathbf{\Lambda}_{\sigma}\right)_{ij} = n_{i} \delta_{ij}$. The creation operators on the NO basis $\tilde{c}_{i\sigma}^{\dagger}$ is expressed in terms of the original operators as $\tilde{c}_{i\sigma}^{\dagger} = \sum_j \left(\bm{U}^{-1} \right)_{ij} c_{j \sigma}^{\dagger}$.
This transformation will mix the impurity and the bath orbitals and lead to complicated interaction term, significantly increasing the storage cost of calculation. Following Ref.~\cite{Lu}, we use a simplified scheme: requiring that the transformation does not mix the impurity and the bath sites. We only need to diagonalize the block-diagonal parts of $\bm{\gamma}_{\sigma}$. Written on the basis $\left\{ d_{1 \sigma}, ..., d_{N_{d} \sigma}, c_{1 \sigma}, ..., c_{N_{b} \sigma}^{\dagger} \right\}$, they read
\begin{eqnarray}  
\label{densitydef}
&& \left( \bm{\gamma}^d_{\sigma}\right)_{ij} = \avstate{\Psi}{d_{i \sigma}^\dagger d_{j \sigma}}{\Psi},    \nonumber \\
&& \left(\bm{\gamma}^c_{\sigma} \right)_{ij} = \avstate{ \Psi}{c_{i \sigma}^\dagger c_{j \sigma} }{\Psi}.
\end{eqnarray}
A block-diagonal unitary matrix $\bm{U}$ is used to diagonalize $\bm{\gamma}^d_{\sigma}$ and $\bm{\gamma}^c_{\sigma}$,
\begin{equation} 
\label{Udef}
\bm{U}_\sigma = 
\left(\begin{array}{cc}
	\bm{U}^d_\sigma 	&	0	\\
	0						&	\bm{U}^c_\sigma
\end{array}\right),
\end{equation}
\begin{eqnarray}  
\label{Uform}
&&	\left( \bm{U}^{d}_{\sigma} \right)^{-1} \bm{\gamma}^{d}_{\sigma} \bm{U}^d_{\sigma} = \bm{\Lambda}^{d},    \nonumber \\
&&		\left( \bm{U}^{c}_{\sigma} \right)^{-1} \bm{\gamma}^{c}_{\sigma} \bm{U}^{c}_{\sigma} = \bm{\Lambda}^{c}.
\end{eqnarray}
Here, $\bm{\Lambda}^{d}$ and $\bm{\Lambda}^{c}$ are diagonal matrices with occupation numbers as the diagonal elements.
In the new basis, the creation operators read
\begin{eqnarray}    
\label{new operator}
&&   \tilde{d}_{i \sigma}^\dagger  = \sum_{j} \left( \bm{U}^{d}_{\sigma} \right)_{ji}^{\ast} d_{j \sigma}^\dagger   \nonumber  \\
&&  \tilde{c}_{i \sigma}^\dagger  = \sum_{j} \left( \bm{U}^{c}_{\sigma} \right)_{ji}^{\ast} c_{j \sigma}^\dagger .
\end{eqnarray} 
The advantage of using the block-diagonal ansatz for $\bm{U}_{\sigma}$ is that in the new operator representation, the Hamiltonian maintains the definition of impurity and bath. As to be shown below, although the new basis is not exact NO basis, it still significantly reduces the number of Slater determinants required by the ground state. For a given $|\Psi \rangle$, finding the above transformation is trivial for systems on the order of $N_{d} \sim 10^{0}$ and $N_{b} \sim 10^{2}$. The ground state itself, however, needs to be refined iteratively by combining the methods such as ED, CI, or Lanczos~\cite{Lin,He1,Lu}.

\begin{figure}[t]
\centering
   \includegraphics[width=0.5\textwidth]{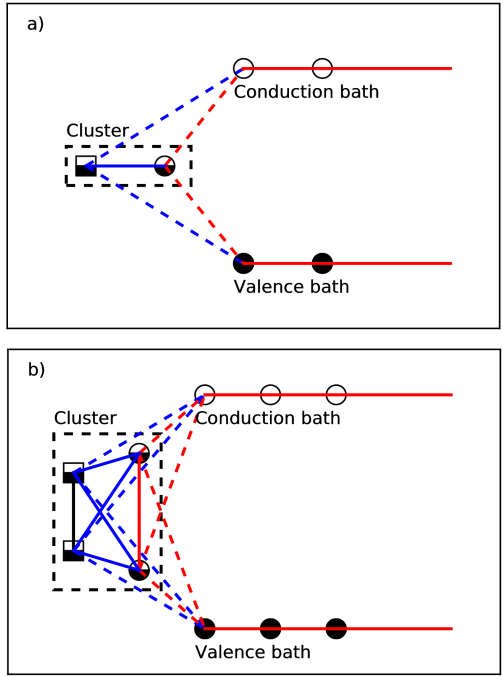}
\caption{(Color online). Pictorial representation of the single-particle orbital basis and the form of Hamiltonian on it. Squares (circles) represent impurity (bath) sites. The electron occupation of each orbital is shown by the degree of filling of the symbol, $n_{i\sigma}=0$ for an empty orbital and $n_{i\sigma}=1$ for an occupied orbital. The red lines, black, and blue lines are for intra-bath, intra-impurity (in case $N_d>1$), and impurity-bath hoppings. The dashed lines show the hoppings between the cluster and the valence/condunction chains. Here, only hoppings to the first condunction/valence sites are shown. \label{Fig_extend_operator}}
\end{figure}

Starting from an approximate ground state, we calculate the density matrices $\bm{\gamma}^{d}_{\sigma}$ and $\bm{\gamma}^{c}_{\sigma}$, diagonalize them, and produce the new operators in Eq.~(\ref{new operator}). The new Hamiltonian after the unitary transformation can be written in terms of $\tilde{d}_{i\sigma}$ and $\tilde{c}_{i\sigma}$ and their Hermitian conjugate. Following Ref.~\cite{Lu}, we introduce a pictorial representation of the new Hamiltonian and the single-particle basis. Fig.~\ref{Fig_extend_operator} shows examples of such figures for a single impurity AIM (Fig.~\ref{Fig_extend_operator}(a)) and for a two-impurity AIM (Fig.~\ref{Fig_extend_operator}(b)).

Fig.~\ref{Fig_extend_operator} (and similar figures in Fig.~\ref{Fig_MG1},\ref{Fig_G_magnetic}, and \ref{Fig_MG2} below) is produced from the following procedure.
(1) We find the natural orbital from the converged ground state;
(2) among the obtained bath orbitals, we identify the valence (filled), conduction (empty), and the partially filled bath orbitals;
(3) write down the Hamiltonian in the new orbital basis; 
(4) tridiagonalize the hopping Hamiltonian of the valance bath and of the conduction bath, separately; and
(5) plot a square for the impurity orbital and a circle for the bath orbital, fill them according to their occupations, and plot a line between every two sites, with the line width proportional to the hopping strength. 
     
Both previous~\cite{Lu} and this study find that the transformed Hamiltonian has the structure shown in Fig.~\ref{Fig_extend_operator}.
First, the bath orbitals fall into three categories, a conduction band that is almost empty, a valence band that is almost fully occupied, and a partially occupied band. The number of partially occupied bath sites equals to that of the impurity sites. The small number of partially filled orbitals is consistent with the fact that the ground state has limited number of Slater determinants on the NO basis, {\it i.e.}, a relatively small active space.
Second, the direct hopping between the valence and the conduction sites are small, as shown in the insets of Fig.~\ref{Fig_MG1}, ~\ref{Fig_G_magnetic}, and ~\ref{Fig_MG2}. When applying the Hamiltonian on a Slater determinant with fully occupied valence and empty conduction orbitals, such hopping terms will generate new Slater determinants with small coefficients only.
Third, after tridiagonalization, the conduction and the valence part of the Hamiltonian can be represented by two separate semi-infinite chains.
Since the tridiagonalization only mixes the NOs with same eigenvalues of the single-particle density matrix, the obtained chain sites still represent NOs. 
 Fourth, if we bound the partially occupied bath sites and the impurity sites into a cluster, they determine the dimension of the active space, or the number of Slater determinants required for a faithful representation of the ground state. 
The hopping strengths between the active orbitals (i.e., cluster orbitals) and the inactive orbitals (i.e., valence and conduction orbitals) are 
determined by the original hopping matrix and the orbital transformation and they have variations in general. From the same argument made for the second point, we expect that the hoppings between the cluster orbitals and the two chains are localized to first few sites of the conduction and valence chains. This is indeed the case in the actual calculation. 

These features of the transformed Hamiltonian guarantee the sparseness of the Hamiltonian matrix and are crucial for the applicability of the NO-based Lanczos algorithm.   
In the schematic picture shown in Fig.~\ref{Fig_extend_operator}, we only plot the hopping from the cluster sites to the first sites of the two chains. Constant line width is used and particle-hole symmetric situation is shown. In the actual calculation shown in the insets of Fig.~\ref{Fig_MG1},\ref{Fig_G_magnetic}, and \ref{Fig_MG2} below, longer range hoppings also exist. The line width varies with sites and the particle-hole asymmetric situation is also considered.

\section{NO-based Lanczos Impurity solver     \label{Impurity solver}} 
\subsection{Lanczos method}
In this subsection, we give a brief overview of the Lanczos approach to the ground state. Details can be found in Ref.~\cite{Dagotto}. For a given initial state $ \rstate{\psi_0} $, the $M$-th order Krylov space is defined as
\begin{equation}
K_M (\rstate{ \psi_0 }) = \Big\lbrace \rstate{ \psi_0 } , \, H \rstate{ \psi_0 },\cdots ,  H^{M-1} \rstate{ \psi_0 } \Big\rbrace,
\end{equation}
where $ K_M(\rstate{\psi_0}) $ is the $M$-dimensional subspace of a $d$-dimensional full Hilbert space. Usually $M \ll d$. A set of orthonormal basis in $K_M (\rstate{ \psi_0 }) $ can be constructed recursively as
\begin{equation} 
\label{Lanczos itration}
\rstate{ \psi_{i+1} } = H  \rstate{ \psi_{i} } - a_{i} \rstate{ \psi_{i} } - b_{i}^2 \rstate{ \psi_{i-1} }, \,\,\,\,\, (i=0,1,\cdots, M-1)
\end{equation}
with the initial values $b_0 \equiv 0$ and $\rstate{\psi_{-1}} \equiv 0$. The coefficients are given by
\begin{eqnarray}
&&  a_{i} = \avstate{ \psi_{i}}{ H }{ \psi_i } / \dotstate{ \psi_i}{ \psi_i} , \nonumber \\
&&  b_{i}^2 = \dotstate{ \psi_{i}}{ \psi_i} / \dotstate{ \psi_{i-1} }{ \psi_{i-1} }.
\end{eqnarray}
After normalization, one obtains the orthonormal Lanczos basis $\lbrace \rstate{\psi_0}, \rstate{\psi_1}, \cdots, \rstate{\psi_{M-1}} \rbrace $ of the subspace $K_M (\rstate{\psi_0}) $. On this basis, the Hamiltonian becomes a tridiagonal matrix
\begin{equation} 
\label{Lanczos tri-matrix}
\bm{T} = 
\left(
\begin{array}{ccccc}
	a_0		&	b_1		& 	{}		&	{}		&	0		\\
	b_1		&	a_1		&	\ddots	&	{}		&	{}		\\
	{}		&	\ddots	& 	\ddots	&	\ddots	&	{}		\\
	{}		&	{}		& 	\ddots	&	a_{M-2}	&	b_{M-1}	\\
	0		&	{}		& 	{}		&	b_{M-1}	&	a_{M-1}
\end{array}
\right)  
\end{equation}
and can be diagonalized by a $M \times M$ unitary matrix $\bf{Q}$ as
\begin{equation}
\bm{D}_T = \bm{Q}^{-1} \bm{T} \bm{Q}.
\end{equation}
The diagonal elements of $\bm{D}_T$ give the approximate eigenvalues $\{ E_m \}$ and the corresponding approximate eigenvectors satisfying $H \rstate{\Psi_m} \approx E_m \rstate{m}$ are given by
\begin{equation}    
\label{Lanczos eigenstate}
\rstate{\Psi_m} = \sum_{i=0}^{M-1} {Q_{im} \rstate{\psi_i}},  \,\,\,\,\,\,\,\,\,\,\,\, (m=1,2, \cdots,  M).
\end{equation}
The convergence of the extremal eigenvalues with increasing Krylov space dimension $M$ is very fast. High precision results can be obtained with $M$ of the order $10^{2}$. The initial state $\rstate{\psi_0}$ can be chosen arbitrary but must have a finite overlap with the ground state.

One could work on a small Krylov space and iterate the process to improve the accuracy of the ground state~\cite{Dagotto}. That is, for the $k+1$-th Lanczos, one can take the ground state of the $k$-th Lanczos calculation as the initial vector, $\rstate{\psi_0}_{k+1} = \rstate{\Psi_1}_{k}$ ($k=1, 2, \cdots$). For the first iteration $k=1$, $\rstate{\psi_0}_1$ is chosen randomly. The iteration stops when the ground energy reaches a given precision,
\begin{equation}
 \label{convergence of Lanczos}
 \big\Vert \, \left( \hat{H}-E_{1k} \right) \, | \Psi_1 \rangle_{k} \, \big\Vert < \epsilon_L .
\end{equation}
In our NO-based Lanczos calculation, we use this iterative Lanczos method with a convergence criterion $\epsilon_L =1.0 \times 10^{-6}$. 
%
%
\subsection{NO-based Lanczos method     \label{NO-based lanczos method}}  

The conventional Lanczos method uses a complete set of Slater determinants as the working basis. Each Lanczos vector $\rstate{\psi_i}$ is stored in the memory as a $d$-dimensional vector and the Hilbert space dimension $d$ increases exponentially with system size, leading to exponential growth of computational cost. As shown in Fig.~\ref{Fig_slater}, the ground state of AIM contains only a tiny fraction of all the Slater determinants if we choose the NO orbital. Those Slater determinants with negligibly small coefficients can be safely ignored. This feature is employed in the NO-based Lanczos method to reduce the memory cost of Lanczos calculation, making it possible to treat AIM with $N_{b} \sim 10^2$~\cite{Lu}. 

We use two iteration loops in our NO-based Lanczos method. The outer loop is the orbital iteration, in which the single-particle density matrix of an approximate ground state is diagonalized to generate the new NO basis. Inside the orbital iteration, there is the sparse Lanczos iteration, in which a new approximate ground state is calculated in a small subspace of Slater determinants on the fixed NO basis. These determinants are picked out by applying the expanding operator to an initial subspace. Below, we describe the two iterations one by one.

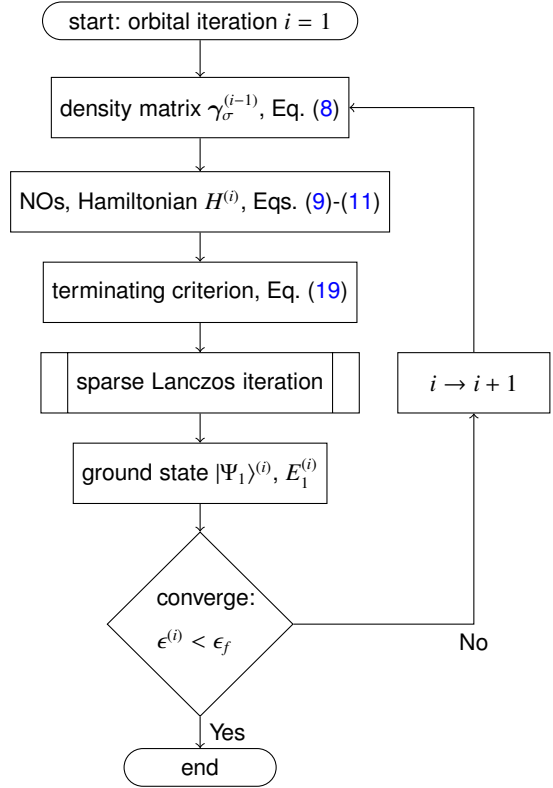
\begin{figure}[t!]
\begin{center}
\begin{tikzpicture}[font={\sf \small}]
ww.
\def \smbwd{2cm}
\thispagestyle{empty}
\node (start) at (0,0) [draw, terminal,minimum width=\smbwd, minimum height=0.5cm] {start: orbital iteration $i=1$};    

\node (density) at (0,-1.15) [draw, process,minimum width=\smbwd, minimum height=0.8cm] {density matrix $\bm{\gamma}^{(i-1)}_\sigma$,  Eq.~(\ref{densitydef}) };      

\node (H) at (0,-2.4) [draw, process, minimum width=\smbwd, minimum height=0.8cm] {NOs, Hamiltonian $H^{(i)}$, Eqs.~(\ref{Udef})-(\ref{new operator})};

\node (epsilon_i) at (0,-3.6) [draw, process, minimum width=\smbwd, minimum height=0.8cm] { terminating criterion, Eq.~(\ref{epsilon_tune}) };

\node (sparse) at (0,-4.8) [draw, predproc, align=left,minimum width=\smbwd,minimum height=0.8cm] {sparse Lanczos iteration};        

\node (GS) at (0,-6.0) [draw, process, minimum width=\smbwd, minimum height=0.8cm] { ground state $\rstate{\Psi_{1}}^{(i)}$, $E_1^{(i)}$  };      

\node (decide) at (0,-8.0) [draw, decision, minimum width=\smbwd, minimum height=0.6cm] {\parbox{3.8em}{ { converge:  $\epsilon^{(i)} < \epsilon_f$} }  };

\node (newepsilon) at (3.6, -4.8) [draw, process, minimum width=\smbwd, minimum height=0.8cm] {  $i \rightarrow i+1$  };

\node (end) at (0,-9.9) [draw, terminal,minimum width=\smbwd,minimum height=0.5cm] {end};

\draw[->] (start) -- (density);
\draw[->] (density) -- (H);
\draw[->] (H) -- (epsilon_i);
\draw[->] (epsilon_i) -- (sparse);
\draw[->] (sparse) -- (GS);
\draw[->] (GS) -- (decide);
\draw[->] (decide) -- node[right]{Yes}(end);
\draw[->] (decide) -| node[below]{No}(newepsilon);
\draw[->] (newepsilon) |- (density);

\end{tikzpicture}
\end{center}
\vspace{0.0in}
\caption{(Color online). Flow chart of the orbital iteration. \label{Fig_floworbital}}
\end{figure}
\subsubsection{Orbital iteration}

The orbital iteration is composed of the following steps.
\begin{enumerate}[1)]

\item For the first iteration $i=1$, we generate the diagonal blocks $\bm{\gamma}^{d(0)}_{\sigma}$ and $\bm{\gamma}^{s(0)}_{\sigma}$ of the density matrix $\bm{\gamma}_{\sigma}^{(0)}$ from an approximate ground state of AIM Eq.~(\ref{AIMH}). One could use the Hartree-Fock approximation or other approximations such as the lattice density functional theory~\cite{LDFT} to produce the approximate ground state. In this paper, we use the Hartree-Fock approximation, {\it i.e.}, $\bm{\gamma}_\sigma^{(0)} = \bm{\gamma}^{HF}_{\sigma}$.

\item \label{orbital iteration begin} For iteration $i \geqslant 1$, diagonalize both the impurity and the bath density matrices $\bm{\gamma}^{d(i-1)}_{\sigma}$ and $\bm{\gamma}^{c(i-1)}_{\sigma}$ to produce $\bm{U}^{(i-1)}_{\sigma}$ according to  Eqs.(\ref{Udef}) and (\ref{Uform}). After the new operators are obtained from Eq.~(\ref{new operator}), $H^{(i)}$ is expressed in terms of the new operators, which has the structure depicted in Fig.~\ref{Fig_extend_operator}.

\item \label{orbital iteration end} The ground state $| \Psi_{1} \rangle^{(i)}$ of $H^{(i)}$ is then solved by the sparse Lanczos iteration.  We set the terminating criterion for the sparse Lanczos iteration as follows,
\begin{eqnarray}
\label{epsilon_tune}
&&   \epsilon^{(i)} = 10^{-4}, \,\,\, \text{or} \,\,\, \, j =\max{(i, 4)}, \,\,\,\,\,\,\,\, \,\, (i \in \text{stage one});   \nonumber \\
&&   \epsilon^{(i)} = \epsilon^{(i-1)}/4, \,\,\,\,\,\,\,\,\,\, (i \in \text{stage two}).  
\end{eqnarray}
That is, we split the orbital iterations into two stages: stage one (the constant-$\epsilon^{(i)}$ stage for small $i$) and stage two (the decreasing-$\epsilon^{(i)}$ stage for large $i$). To get the maximum efficiency, we use different strategies in these two stages to terminate the Lanczos iteration. 
The orbital iteration terminates when $\epsilon^{(i)} < \epsilon_f = 3 \times 10^{-7}$. 

\item If $\epsilon^{(i)} > \epsilon_f$, calculate the new density matrix $\bm{\gamma}^{(i)}_\sigma$ from $\rstate{\Psi_1}^{(i)}$ using Eq.~(\ref{densitydef}). Go back to step \ref{orbital iteration begin}).
 
\end{enumerate}

\subsubsection{Sparse Lanczos iteration}

After the $i$-th orbital transformation, the approximate NO basis and $H^{(i)}$ are obtained. For the fixed NO basis and $H^{(i)}$, we call the sparse Lanczos process to calculate the ground state and energy of $H^{(i)}$ to a given precision $\epsilon^{(i)}$. We use the following iterative scheme to avoid using the complete working basis. We start from a small subspace containing a few most important Slater determinants. We then expand the subspace by adding to it some new Slater determinants which are generated by applying the expanding operator on the subspace. Lanczos calculation is done in the expanded subspace to produce the ground state of $H^{(i)}$. From this ground state, we find out those Slater determinants that are unimportant and remove them from the subspace, compressing the subspace. We finally obtain a new subspace which is in general larger than the original one but is more relevant to the true ground state. We start the next round of expanding, Lanczos, and compressing process. This iteration is carried on until the ground state reaches a prescribed precision $\epsilon^{(i)}$.

For the expanding process, one needs to enlarge the subspace for Lanczos in such a way that, first, only important Slater determinants are added and, second, the dimension of the subspace increases in a controlled way so that the subsequent Lanczos calculation can be carried out without too much resources. If the NO-based Lanczos method works in practice, as the orbital iteration carries on, the maximum dimension of the subspace after compressing should saturate to a constant value, which is close to the actual number of Slater determinants required to describe the ground state to a given precision. For AIM, we find that the saturated subspace dimension is on the order of $10^{3}$ for $\epsilon_f =10^{-7}$ which is sufficient for most purposes. In Ref.~\cite{Lu}, $\epsilon_f =10^{-14}$ is used for systems with $N_{b}=301$ and the saturated subspace dimension is $10^{9}$.

Suppose we have an initial subspace $S^0$ which contains a given set of Slater determinants, $S^{0} = \lbrace \rstate{w_1}, \rstate{w_2}, \cdots, \rstate{w_{n0} } \rbrace$. In the $i$-th orbital iteration, the expanded subspace $S^{1}$ reads
\begin{equation}
S^1 = S^0 \cup \hat{C}^{(i)} S^0. 
\end{equation}
In general, one could apply $\hat{C}^{(i)}$ to $S^0$ $n$ times to generate a large enough subspace $S^n$. The simplest $\hat{C}^{(i)}$ is the linear combination of density operators, such as $\hat{C}^{(i)} = \sum_{\sigma} \left( c_{1\sigma}^{\dagger} c_{2 \sigma} + h.c. \right) + \left( d_{1 \sigma}^{\dagger} c_{2 \sigma} + h.c. \right)$. Here $d_{i \sigma}$ and $c_{j \sigma}$ are respectively the annihilation operators of the impurity and the bath NOs in the $i$-th orbital iteration. $C^{(i)}$ should include all the hopping terms of the Hamiltonian. In this paper, we first include in $C^{(i)}$ all the intra-cluster and intra-valence/conduction band hoppings. To prevent the space from increasing too fast, for each cluster site, among its hoppings to all the valence/conduction sites, we only keep the largest one. We also neglect the valence-conduction hopping terms in $C^{(i)}$. Note that once the subspace is generated, we use the full $H^{(i)}$ without approximation in the Lanczos calculation to get the ground state. The pictorial representation of $\hat{C}^{(i)}$ is the same as $H^{(i)}$ and is given in Fig.~\ref{Fig_extend_operator}, both for the single- and for the two-impurity AIMs. 

\begin{figure}[t]
\centering
   \includegraphics[width=0.8\textwidth]{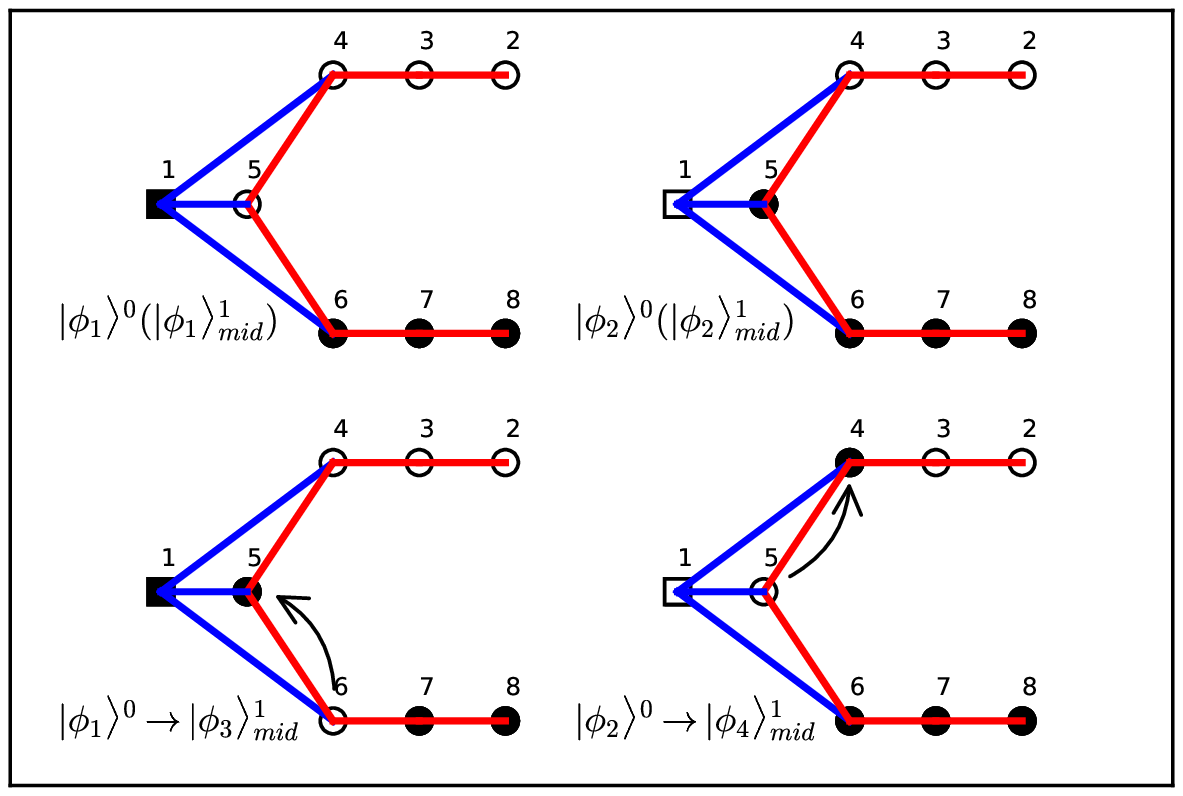}
\caption{(Color online). The subspace $S^0$ and $S^1_{mid}$ for single Anderson model containing $7$ bath, neglecting the spin indices. $S^0 = \lbrace \rstate{\phi_1}^0 , \rstate{\phi_2}^0 \rbrace, S^1_{mid} = \lbrace \rstate{\phi_1}^0 $, $\rstate{\phi_2}^0,\rstate{\phi_3}^1_{mid} , \rstate{\phi_4}^1_{mid} \rbrace$. \label{Fig_extend_iteration}}
\end{figure}

\begin{figure}[t!]
\begin{center}
\begin{tikzpicture}[font={\sf \small}]

\def \smbwd{2cm}
\thispagestyle{empty}
\node (start) at (0,0) [draw, terminal,minimum width=\smbwd, minimum height=0.5cm] {start: for given NOs, $H^{(i)}$, and $\epsilon^{(i)}$  };    

\node (iniLanc) at (0, -1.1) [draw, process,minimum width=\smbwd, minimum height=0.8cm] { prepare $S^{0}$, set Lanczos index $j=1$ };    

\node (initialspace) at (0,-2.4) [draw, process,minimum width=\smbwd, minimum height=0.8cm] { expand $S^{j-1}$ to $S^{j}_{\text{mid}}$, Eq.~(\ref{extendS}) };      

\node (extend) at (0,-3.6) [draw, process, minimum width=\smbwd, minimum height=0.8cm] {set initial Lanczos vector $|\psi_0 \rangle^{j}$, Eq.(\ref{init_psi0})  };

\node (Lanczos) at (0,-4.8) [draw, predproc, align=left,minimum width=\smbwd,minimum height=0.8cm] {iterative Lanczos in $S^{j}_{\text{mid}}$ up to $\epsilon_L$};        

\node (GS) at (0,-6.0) [draw, process, minimum width=\smbwd, minimum height=0.8cm] {ground state $\rstate{\Psi_1}^{j}_{\text{mid}}$, energy $E_{1}^{j}$};      

\node (reduce) at (0,-7.2) [draw, process, minimum width=\smbwd, minimum height=0.8cm] {compress: $S^{j}_{\text{mid}} \rightarrow S^{j}$, $| \Psi_1 \rangle^{j}_{mid} \rightarrow | \Psi_1 \rangle^{j}$, Eq.~(\ref{compressspace})};      

\node (updateLanc) at (3.7,-6.0) [draw, process, minimum width=\smbwd, minimum height=0.8cm] {  $j \rightarrow j+1$  };      

\node (decide) at (0, -9.2) [draw, decision, minimum width=\smbwd, minimum height=0.6cm] {\parbox{3.8em}{converge: Eq.~(\ref{convergence_lanczos}) } };

\node (output) at (0,-11.2) [draw, process, minimum width=\smbwd,minimum height=0.5cm] {output $| \Psi_1 \rangle^{(i)}$, $E_{1}^{(i)}$, $S^{(i)}$, Eq.~(\ref{output_Lanczos}) };

\node (end) at (0,-12.1) [draw, terminal, minimum width=\smbwd,minimum height=0.5cm] {end};

\draw[->] (start) -- (iniLanc);
\draw[->] (iniLanc) -- (initialspace);
\draw[->] (initialspace) -- (extend);
\draw[->] (extend) -- (Lanczos);
\draw[->] (Lanczos) -- (GS);
\draw[->] (GS) -- (reduce);
\draw[->] (reduce) -- (decide);
\draw[->] (decide) -- node[right]{Yes}(output);
\draw[->] (decide) -| node[below]{No}(updateLanc);
\draw[->] (updateLanc) |- (initialspace);
\draw[->] (output) -- (end);
\end{tikzpicture}

\end{center}
\vspace{0.0in}
\caption{(Color online). Flow chart of the sparse Lanczos iteration. \label{Fig_flowparse}}
\end{figure}
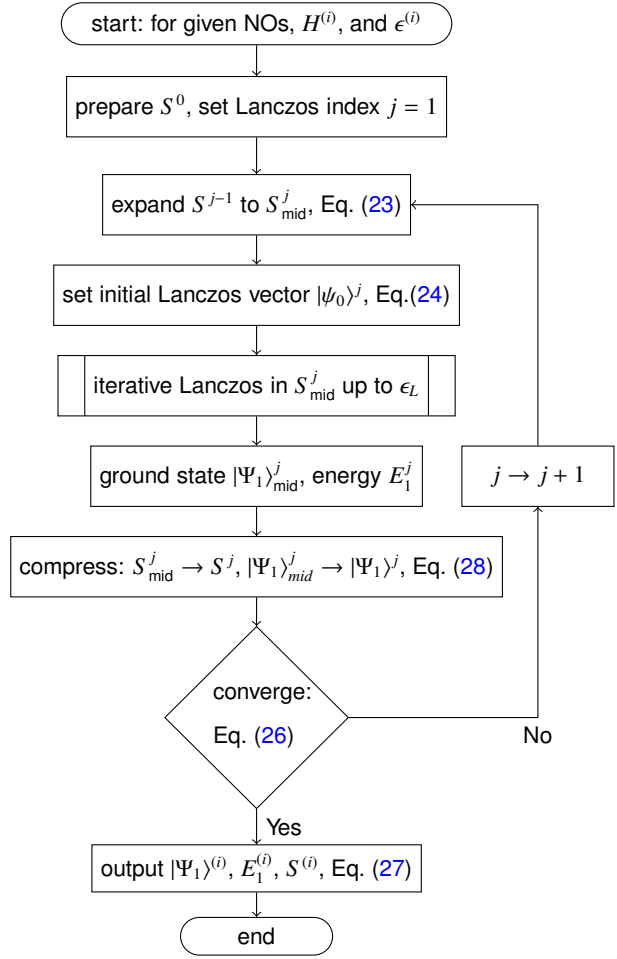

In the following, we describe the algorithm of sparse Lanczos. For simplicity, we only discuss the single-impurity AIM and neglect the spin indices. Here the $i$ is used to denote the orbital iteration number, and $j$ to denote the Lanczos iteration number.

\begin{enumerate}[1)]

\item \label{initial space}  Inside the $i$-th orbital iteration where both NO orbital and $H^{(i)}$ are fixed, choose an initial subspace $S^{0}$. For the first several orbital iterations (small $i$), we use the ground state at $\mu=U=0$ to construct the NOs. Since we diagonalize the impurity and the bath density matrices separately, in the particle-hole symmetric case, we obtain two NOs and form the subspace $S^{0}$ as shown in Fig.~\ref{Fig_extend_iteration} (with spin indices neglected), 
\begin{equation} 
\label{initialS}
S^{0} =  \lbrace \rstate{\phi_1}^{0} , \rstate{\phi_2}^{0} \rbrace.
\end{equation}
In the particle-hole asymmetric case, $S^{0}$ contains all the $4^2=16$ different configurations of the two cluster sites. 
Once the NOs become stable (larger $i$), we use 
\begin{equation}   
\label{initialS2}
  S^{0} =  \lbrace \rstate{\phi_1}^{0} , \rstate{\phi_2}^{0} \rbrace \cup S^{(i-1)},
\end{equation}
where $S^{(i-1)}$ is the subspace of Slater determinants appearing in the ground state of $H^{(i-1)}$ with updated NOs. Eq.(\ref{initialS2}) guarantees that the important Slater determinants $\rstate{\phi_1}^{0}$ and $\rstate{\phi_2}^{0}$ are always included in $S^{0}$. 

\item Construct the expanding operator $\hat{C}^{(i)}$ from the dominant non-interacting part of $H^{(i)}$, as described in the above text and pictorially shown in Fig.~\ref{Fig_extend_operator} (a).

\item \label{expand} For $j=1,2,...$, expand the subspace $S^{j-1}$ to obtain $S^{j}_{\text{mid}}$,
\begin{equation} 
\label{extendS}
S^{j}_{\text{mid}} = S^{j-1}  \cup \hat{C}^{(i)} S^{j-1}, \,\,\,\,\,\,\,\,\ (j= 1,2,\cdots).
\end{equation}
For an example, $S^{1}_{\text{mid}}$ contains four Slater determinants which are shown pictorially in Fig.~\ref{Fig_extend_iteration}.

\item \label{sparse begin} In the subspace $S^{j}_{\text{mid}} = \lbrace \rstate{\phi_1}^{j}, \rstate{\phi_2}^{j}, \cdots, \rstate{\phi_{n_{\text{mid}}}}^{j} \rbrace$, do iterative Lanczos calculation for $H^{(i)}$ to produce the ground state $\rstate{\Psi_{1}}^{j}_{\text{mid}}$ and energy $E_{1}^{j}$ to a given precision as described by Eq.~(\ref{convergence of Lanczos}). We use the Krylov space dimension $M=40$. 
For the initial Lanczos vector $| \psi_0 \rangle^{j}$, we use the following scheme,
\begin{eqnarray}
\label{init_psi0}
   && | \psi_0 \rangle^{j} = | \psi \rangle_{rand}, \,\, \,\,\, \,\,\, \,\,\, (j=1, i \in \text{stage one}); \nonumber \\
   && | \psi_0 \rangle^{j} = | \Psi_1 \rangle^{(i-1)}, \,\, \,\,\, \,\,\, \,\,\, (j=1, i \in \text{stage two} ); \nonumber \\   
   && | \psi_0 \rangle^{j} = | \Psi_1 \rangle^{j-1},  \,\, \,\,\,\,\,\, \,\,\, (j \geqslant 2).
\end{eqnarray}      
Here $| \psi \rangle_{rand}$ is a random vector in subspace $S^{0}$. The two stages of the orbital iteration are defined below Eq.({\ref{epsilon_tune}} ).
The ground state and energy are obtained as 
\begin{eqnarray}   
\label{eigstates}
&& \rstate{\Psi_{1}}^{j}_{\text{mid}} = \sum_{k=1}^{n_{\text{mid}}}{ \alpha_k \rstate{\phi_{k}}^{j} },    \nonumber  \\
&& E_{1}^{j} = {}_{\text{mid}}^{\,\,\,\,\,\,\,j}\avstate{\Psi_{1}}{H^{(i)}}{\Psi_{1}}^{j}_{\text{mid}}.
\end{eqnarray}
The iteration is terminated if
\begin{equation}    
\label{convergence_lanczos}
 {\Big|} \, {}_{\text{mid}}^{\,\,\,\,\,\,\,j} \langle \Psi_{1} \big| [ H^{(i)} ]^2 \big| \Psi_{1} \rangle^{j}_{\text{mid}} - \left[ {}_{\text{mid}}^{\,\,\,\,\,\,\,j} \langle \Psi_{1} \big|  H^{(i)} \big| \Psi_{1} \rangle_{\text{mid}}^{j} \right]^2  \, {\Big|} < \epsilon^{(i)} \big| E_{1}^{j} \big|^2.
\end{equation}
 This criterion measures how close $E_{1}^{j}$ is to the true  ground state energy of $H^{(i)}$ in the full Hilbert space, being different from the subspace precision $\epsilon_L$ in Eq.(\ref{convergence of Lanczos}). 

The ground state and the final subspace are given as  
\begin{eqnarray}  \label{output_Lanczos}
 &&  \rstate{\Psi_{1}}^{(i)} = \rstate{\Psi_{1}}^{j}_{\text{mid}}, \nonumber \\
 &&  E_{1}^{(i)} = E_{1}^{j} ,  \nonumber \\
 &&  S^{(i)} = S^{j}_{\text{mid}}.
\end{eqnarray}
Otherwise, do step \ref{remove unimportant determinants}). 

\item \label{remove unimportant determinants} To maintain a relatively small subspace, some determinants with small coefficients $\alpha_k$ in the ground state Eq.~(\ref{eigstates}) will be removed from the subspace $S^{j}_{\text{mid}}$. The compressed subspace is denoted as $S^{j}$ with dimension $n_j$, and the ground state in this subspace $\rstate{\Psi_{1}}^{j}$, 
\begin{eqnarray} 
\label{compressspace}
&&  S^{j} =  \left\lbrace \rstate{\phi_k}^{j}, \,\,\,\,(k=1,2,...,n_j) \,\,\,\, \Big\vert  \,\, \,\,  \rstate{\phi_k}^{j} \in S^{j}_{\text{mid}}, \,\,\,\, \alpha_k^2 \geqslant \epsilon_{\text{cut}}  \right\rbrace ,  \nonumber \\
&& \rstate{\Psi_{1}}^{j} =  \frac{1}{c} \sum_{k=1}^{n_j} {\alpha_k \rstate{\phi_{k}}^{j}}.
\end{eqnarray}
Here $c = \sqrt{\sum_{k=1}^{n_j} \alpha_k^{2}}$ is the normalization constant. We use a cut-off value $\epsilon_{\text{cut}} = \epsilon^{(i)}$.
Go back to Step \ref{expand}) with the initial Lanczos vector $\rstate{\psi_0}^{j+1} = \rstate{\Psi_{1}}^{j}$. This process is iterated until the required precision is met.
\end{enumerate}

\begin{figure}[t]
\vspace{-2cm}
\centering
   \includegraphics[width=0.8\textwidth]{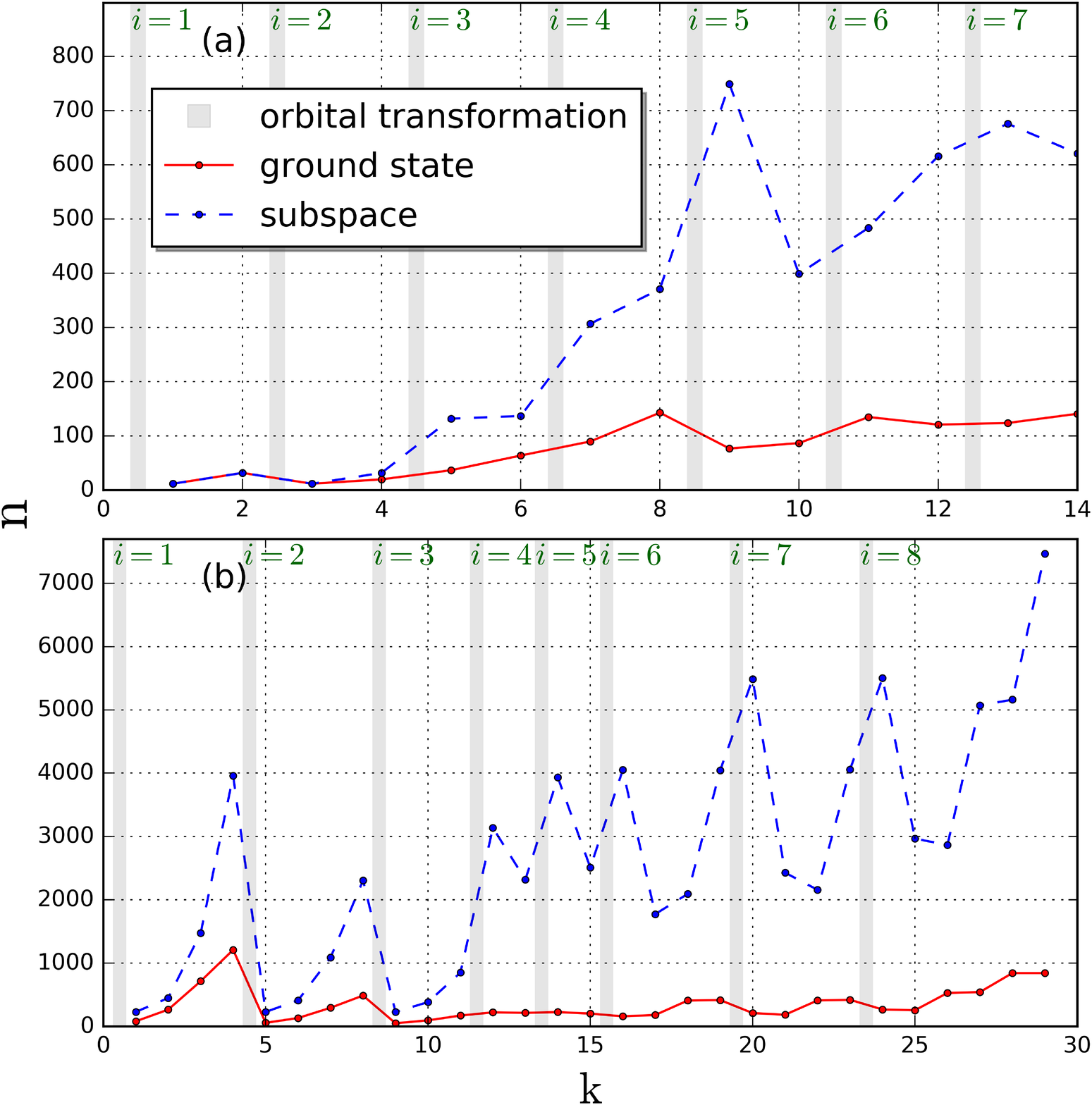}
\caption{(Color online). The number of Slater determinants $n(k)$ in the ground state after the compressing (solid red line) and in the subspace after expansion (dashed blue line). $k$ is the accumulated Lanczos iteration number in all orbital iterations. The grey bars mark the orbital transformations and dots on the curves mark the sparse Lanczos calculation. (a) Single impurity AIM with $27$ bath sites, $\mu=10 \pi \Delta$, $U = 2 \mu$. The $\epsilon^{(i)}$ values for $i=1 \sim 7$ are respectively $1.00 \times 10^{-4}$, $1.00 \times 10^{-4}$, $1.00 \times 10^{-4}$, $2.50 \times 10^{-5}$, $6.25 \times 10^{-6}$, $1.56 \times 10^{-6}$, and $3.91 \times 10^{-7}$.
(b) two-impurity AIM with $10$ bath sites, $\mu_1=10 \pi \Delta$, $U_1 = 2 \mu$, $\mu_2= 20 \pi \Delta$, $U_2 = 2 \mu_2$, and $U_{12} =15 \pi \Delta$. $\pi \Delta =0.02$. The $\epsilon^{(i)}$ values for $i=1 \sim 8$ are respectively $1.00 \times 10^{-4}$, $1.00 \times 10^{-4}$, $1.00 \times 10^{-4}$, $1.00 \times 10^{-4}$, $2.50 \times 10^{-5}$, $6.25 \times 10^{-6}$, $1.56 \times 10^{-6}$, and $3.91 \times 10^{-7}$.
\label{Fig_slater-iteration}  }
\end{figure}

\subsubsection{Performance analysis}


The amounts of Slater determinants $n$ involved in the calculation are shown as functions of the iteration number $k$ in Fig.~\ref{Fig_slater-iteration}. The data for the single impurity AIM with $N_{b} =27$ bath sites are shown in Fig.~\ref{Fig_slater-iteration}(a) and those for the two-impurity AIM with $N_{b}=10$ in Fig.~\ref{Fig_slater-iteration}(b). For both models, the involved number of Slater determinants saturates in the large iteration regime where $\epsilon^{(i)}$ decreases to $10^{-7}$. For the single-impurity AIM, this maximum subspace dimension is less than $800$ and the ground state has less than $200$ Slater determinants at the precision $10^{-7}$. These numbers for the two-impurity AIM are $8000$ and $1000$, respectively. 


In our calculation, in order to accelerate the convergence and to avoid too fast increase of the subspace dimension, besides the standard algorithm stated above, we used some tricks to optimize the calculation. We split the iteration process to two stages. In the first stage, we fix the error $\epsilon^{(i)}=10^{-4}$ for each sparse Lanczos iteration. The sparse Lanczos iteration is terminated either when this precision is reached or when the iteration number $j =  max(i, 4)$ is reached.
This is done because in the first several orbital iterations, the NOs are still inaccurate and it is meaningless to do the Lanczos with extremely high precision. Instead, we increase the precision of the Lanczos calculation gradually as the quality of NO is improved with the orbital iteration. Choosing $j= max(i, 4)$ is a convenient trick for this purpose. In this stage, we use a random vector in the subspace $S^{0} =  \lbrace \rstate{\phi_1}^{0} , \rstate{\phi_2}^{0} \rbrace$ as the initial Lanczos vector. Correspondingly, in Fig.~\ref{Fig_slater-iteration}(a) and (b), dips appear after at each orbital transformation (grey bar), such as $k=3$ in Fig.~\ref{Fig_slater-iteration}(a) and $k=5$ and $9$ in Fig.~\ref{Fig_slater-iteration}(b). This is because we use $S^0$ as the initial space for every sparse Lanczos iteration. 
When we observe the real convergence of Lanczos calculation Eq.~(\ref{convergence_lanczos}), we go to the next stage. 

In the second stage, when the orbital iteration enters a stable track, for each new orbital basis, we use Eq.~(\ref{initialS2}) to prepare the initial subspace $S^{0}$ for expansion. We also use the ground state of last orbital iteration with updated orbitals as the initial Lanczos vector $\rstate{\psi_{0}}$. In this stage, the dimension of the subspace increases with the Lanczos iteration and reaches a peak value at each orbital transformation. This is because the ground state in the previous orbital basis contains some Slater determinants that are unimportant in the new orbital basis and are removed in the compression process. The peak value of the subspace dimension saturates as the transformation matrix $\bm{U}_{\sigma}$ approaches unity after about $10 \sim 20$ total iterations.

For the single impurity case shown in Fig.~\ref{Fig_slater-iteration}(a), at the first stage ($i=1$, $2$, $3$), the terminating condition $j=max(i, 4)$ for the Lanczos iteration has not been met before the precision $10^{-4}$ is reached after two Lanczos iterations, giving two dots inside neighbouring grey bars. From $i=4$ on, the orbital iteration enters the second stage and the criterion becomes $\epsilon^{(i)} = \epsilon^{(i-1)}/4$.
  In Fig.~\ref{Fig_slater-iteration}(b) for the two-impurity case, we find that in the orbital iteration $i = 1$ and $i= 2$, the precision $10^{-4}$ is not reached before $j=max(i,4) = 4$
 is satisfied, giving $4$ dots inside neighboring grey bars. For $i=3$ and $i=4$, the precision $10^{-4}$ is reached first. The number of Lanczos iterations are $3$ and $2$, respectively. From $i=5$ on, the orbital iteration enters the second stage and the criterion becomes $\epsilon^{(i)} = \epsilon^{(i-1)}/4$.

\begin{figure}[t]
\vspace{-2cm}
   \centering
   \includegraphics[width=0.8\textwidth]{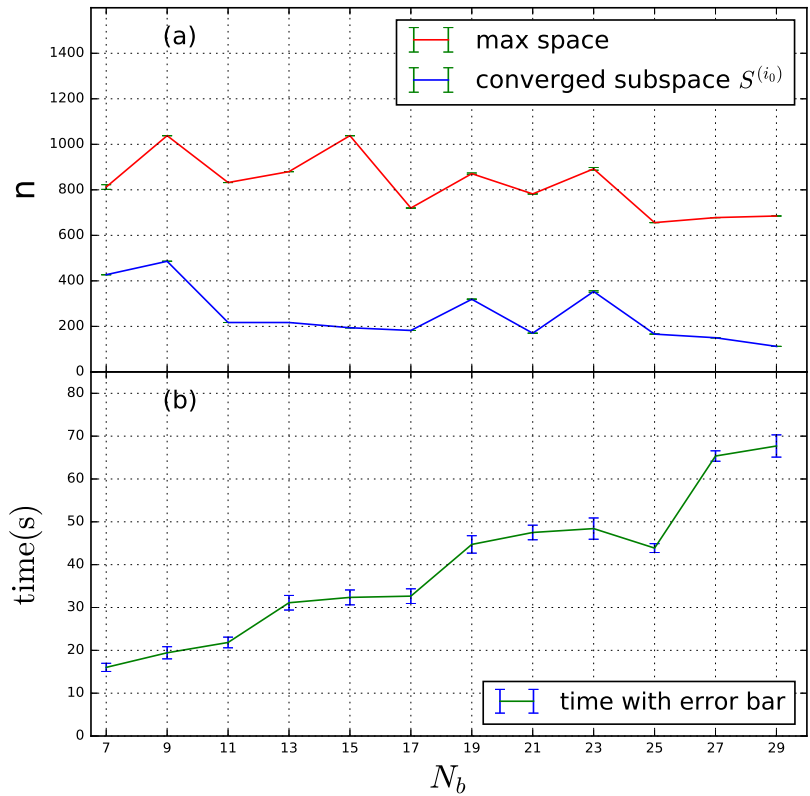}
\caption{(Color online). (a) The maximum subspace dimension (red line) and the number of Slater determinants in the ground state (blue) as functions of $N_{b}$. (b) The computation time for the ground state (green line) as a function of $N_{b}$. The data is from the machine with one $2.80$ GHz Intel(R) Xeon(R) CPU. The results slightly depend on the bath parameters of AIM. The error bars are obtained from results of $10$ calculations with different fittings of the same hybridization function. 
\label{Fig_slaterBath}}
\end{figure}

In Fig.~\ref{Fig_slaterBath}, we show the amount of involved Slater determinants and calculation time as functions of $N_{b}$ for the ground state of single-impurity AIM. In Fig.~\ref{Fig_slaterBath}(a), the number of maximum subspace dimension and the number of Slater determinants involved in the ground state are shown. Both decrease weakly as $N_{b}$ increases and their ratio weakly increases from $3$ to $10$. Only about $10\%$ of the Slater determinants in our subspace finally appears in the ground states. This shows great potential of the present algorithm to handle a large $N_{b}$. This is also demonstrated in previous works~\cite{Lu,He1}.

The weakly decreasing behavior in Fig.\ref{Fig_slaterBath}(a) deserves some discussion. The ground state of AIM is a linear combination of a large number of Slater determinants. On the NO basis, only hundreds of them have significant weights in the ground state and their number is almost independent of $N_b$, if we do not consider the shift of the bath energy levels with $N_b$. As $N_b$ increases, a better fit of the hybridization gives denser bath energy levels close to the Fermi energy, which have smaller influences on the ground state energy and generate more insignificant Slater determinants than those generated by the enlargement of Hilbert space. Under a fixed precision, this leads to the weakly decrease of the number of Slater determinants in the ground state and in the subspace.

In Fig.~\ref{Fig_slaterBath}(b), the calculation time for the ground state of single impurity AIM is linear in $N_{b}$. For the single impurity AIM with $N_{b}=29$, it takes about one minutes on our machine. In the present algorithm, the most time-consuming operations are (a) applying $C^{(i)}$ to a space of Slater determinants, and (b) calculating $H^{(i)} \psi$ in the Lanczos iteration. The number of hopping terms contained in $C^{(i)}$ is proportional to $N_b$ due to the neglect of valence-conduction hoppings. However, the hopping terms in $H^{(i)}$ are conserved and their number increases as $N_{b}^{2}$ in the large $N_b$ limit, with a small coefficient. Therefore, we expect that the observed linear $N_b$ dependence of computing time will finally change into $N_{b}^2$ for larger $N_b$.

It should be noted that although our algorithm is efficient for calculating the ground state of single- and multiple-impurity Anderson models, and for the Green's function of a single-impurity Anderson model, the calculation of Green's function for two-impurity Anderson model is significantly slow. In this work, we use the standard Lanczos method to calculate Green's functions (see below) and confine our demonstration for the two-impurity Green's functions to $N_b=14$.
There are other methods to calculate Green's functions~\cite{Kuehner}. It is the future work to explore whether these methods can produce the Greens' function more efficiently for the multiple impurity cases.

\section{Green's Function   \label{Green's Function}} 
\subsection{Lanczos for zero temperature GF}
In this section, we show how to calculate the zero-temperature single-particle retarded GF $G^{r}_{\sigma}(\omega)$ after the ground state $\rstate{\Psi_g}$ and the energy $E_{g}$ of AIM are obtained using the above NO-based Lanczos method. The retarded GF considered here is defined as $G^{r}_{\sigma}(\omega) = \int_{-\infty}^{\infty} G^{r}_{\sigma}(t-t^{\prime}) \exp{ \left[ i (\omega+ i \eta) (t-t^{\prime}) \right]} d(t-t^{\prime})$ and
\begin{equation}
  G_{\sigma}^{r}(t-t^{\prime}) = \frac{1}{i\hbar} \theta(t-t^{\prime}) \langle \{ d_{\sigma}(t), d_{\sigma}^{\dagger}(t^{\prime}) \ \} \rangle.
\end{equation}
Here $\{A, B\}$ is the anti-commutator of two operators $A$ and $B$. The average $\langle ... \rangle$ is the ground state average of $H$. $\eta$ is an infinitesimal positive number. For a non-degenerate ground state $\rstate{\Psi_g}$ (with energy $E_{g}$), $G^{r}_{\sigma}(\omega)$ can be written into
\begin{eqnarray}    \label{Green expression}
G^{r}_{\sigma}(\omega) &=& G^{>}(\omega)+ G^{<}(\omega),  \nonumber \\
G^{>}(\omega) &=& \avstate{ \Psi_{g}} 
{d_{\sigma}  \frac{1}{\omega + i \eta + E_{g} - \hat{H} }  d^{\dagger}_{\sigma}}
{\Psi_{g}},      \nonumber \\
G^{<}(\omega) &=& \avstate{ \Psi_{g}} 
{ d^{\dagger}_{\sigma}  \frac{1}{\omega + i \eta - E_{g} + \hat{H} } d_{\sigma}} 
{\Psi_{g}}.
\end{eqnarray}
For degenerate ground states, the ground state average is replaced by the average over all ground states. In our calculation we take $\eta = 0.02$.

Following the standard procedure~\cite{Dagotto}, from Eq.~(\ref{Green expression}), the contribution from the particle excitations $G^{>}(\omega)$ can be regarded as the $(1,1)$ element of the matrix $\langle \Psi_g | d_{\sigma} d_{\sigma}^{\dagger} | \Psi_g \rangle \left[ (\omega + i \eta + E_{g}) \bm{1} - \bm{H} \right]^{-1}$, for which the first basis state is chosen as $d_{\sigma}^{\dagger} \rstate{\Psi_g} /\langle \Psi_g | d_{\sigma}d_{\sigma}^{\dagger} | \Psi_g \rangle^{1/2}$. 
Starting a second Lanczos calculation similar to Eqs.~(\ref{Lanczos itration})-(\ref{Lanczos tri-matrix}), but with the first Lanczos vector chosen as $\rstate{\psi_0} = d_{\sigma}^{\dagger} \rstate{\Psi_g}$, one gets the tridiagonal form of $\bm{H}$ on the normalized Lanczos vectors as
\begin{equation}   
\label{Hmat}
\bm{H} = 
\left(
\begin{array}{ccccc}
	a_0		&	b_1		& 	{}		&	{}		&	0		\\
	b_1		&	a_1		&	\ddots	&	{}		&	{}		\\
	{}		&	\ddots	& 	\ddots	&	\ddots	&	{}		\\
	{}		&	{}		& 	\ddots	&	a_{K-2}	&	b_{K-1}	\\
	0		&	{}		& 	{}		&	b_{K-1}	&	a_{K-1}
\end{array}
\right)  
\end{equation}
Here $K$ is the dimension of the Krylov space. Eq.~(\ref{Hmat}) is then diagonalized and the zero-temperature GF is calculated using Lehmann representation as usual. One could also use the continued fraction formula for the $(1,1)$ element of an inverse tridiagonal matrix~\cite{Dagotto,Lu}.

\subsection{NO-based Lanczos for GF}
In this part, we adapt the general Lanczos method described above to the NO-based Lanczos method for calculating GF. In the NO-based Lanczos method, after the convergence, both the Hamiltonian of AIM $H^{(i_0)}$ and the ground state $\rstate{\Psi_1}^{(i_0)}$ are expressed on the converged NO basis. Here the superscript $i_0$ denotes quantities obtained after the last orbital iteration. On this basis, the impurity operators $d_{\sigma}$ and $d_{\sigma}^{\dagger}$ becomes $\tilde{d}_{\sigma}$ and $\tilde{d}_{\sigma}^{\dagger}$, respectively. 

Starting from $\rstate{\psi_0} = \tilde{d}_{\sigma}^{\dagger} \rstate{\Psi_1}^{(i_0)}$ or $\rstate{\psi_0} = \tilde{d}_{\sigma} \rstate{\Psi_1}^{(i_0)}$, the second round of Lanczos procedure involves repeated acting of $H^{(i_0)}$ on $\rstate{\psi_0}$. Although $\rstate{\Psi_1}^{(i_0)}$ contains a limited number of Slater determinants, the number of determinants generated in this process still increases exponentially with $K$, the Krylov space dimension. Therefore, we introduce a truncation of the space. Our algorithm is described below for $G^{>}(\omega)$ and similar procedure is applied to $G^{<}(\omega)$.

\begin{enumerate}[1)]
\item Calculate the ground state $\rstate{\Psi_{g}}$ using NO-based Lanczos. We denote the converged subspace, Hamiltonian, and the ground state as $S^{(i_0)}$, $H^{(i_0)}$, and $\rstate{\Psi_1}^{(i_0)}$, respectively. Associated with $H^{(i_0)}$ is an expanding operator $\hat{C}^{(i_0)}$ same as in the ground-state calculation. Note that each of the hopping terms in $\hat{C}^{(i_0)}$ can generate a new Slater determinant which will be added into the expanded subspace as a separate basis state.

\item Using operator $\tilde{d}^\dagger_{\sigma}$ to expand the subspace, $S^{d+} = \tilde{d}^\dagger_{\sigma} S^{(i_0)}$. The initial Lanczos vector is set up in this space as $\rstate{\psi_0} = \tilde{d}^\dagger_{\sigma} \rstate{\Psi_{1}}^{(i_0)}$.

\item Construct the working space by $S^{K+} = \sum_{j=0}^{K-1} \left[ \hat{C}^{(i_0)} \right]^{j} S^{d+}$. The  amount of determinants in $S^{K+}$ increases rapidly with $K$. To control the computational complexity, we stop extending $S^{d+}$ when the number of determinants is greater than $n_{cut}$. Here we use $n_{cut} = 30000$. 

\item In the subspace $S^{K+}$, calculate $G^{>}(\omega)$ using Lanczos method.

\item The contribution from hole excitations $G^{<} (\omega)$ is calculated similarly in the subspace $S^{K-}$. Here the working subspace $S^{K-} = \sum_{j=0}^{K-1} { \left[\hat{C}^{(i_0)} \right]^{j} S^{d-}}$ and $S^{d-} = \tilde{d}_{\sigma} S^{(i_0)}$.

\end{enumerate}

Before presenting the numerical results, it is interesting to compare the above NO-based Lanczos method with the variational exact diagonalization (VED) method in Ref.~\cite{Schuler}. Both methods use the cluster-plus-bath scheme to represent the transformed/auxilliary Hamiltonian. The cluster contains all the impurity orbitals plus one bath site for each impurity orbital. In the implementation of both methods, the unitary transformation for the impurity orbital is neglected. There are also prominent differences. The VED is a variational scheme which mapps the original Anderson model to an auxilliary one using the Peierls-Feynman-Bogoliubov variational principle and solves the latter using exact diagonalization. NO-based Lanczos is not a variational method and it cannot provide the upper-bound of the ground state energy. 
NO-Lanczos is a method for ground state only while VED applies to all temperatures. In terms of the number of Slater determinants involved in the calculation, VED uses exactly $4^{2 N_d}$ Slater determinants to describe the eigenstate of a $N_d$-orbital Anderson model. This number in NO-based Lanczos is determined by the precision one needs to achieve and is usually larger than $4^{2 N_d}$ since the residual coupling to the rest of the bath are also taken into account.


\section{Results     \label{results}} 
In this section, we solve the single- as well as the two-impurity AIMs with Lorentzian hybridization function Eq.~(\ref{hybrid Lorentzian}) using the NO-based Lanczos method. We compare the GFs with the results from existing numerical methods. The spectral functions are compared with those from exact Lanczos method for small $N_b$, and with the numerical renormalization group (NRG) results for a continuous bath. The NRG results are obtained from the full-density matrix NRG algorithm~\cite{Weichselbaum} with the self-energy trick~\cite{Bulla}, averaged over $8$ interleaved discretizations~\cite{Yoshida}, using $\Lambda=1.6$ and keeping $M_s=256$ states.

The Matsubara GF is compared to the CT-QMC results for a continuous bath. The comparison are made for single- and two-impurity AIMs, including the situations of weak/strong interactions, and with/without particle-hole symmetries. The CT-QMC results are obtained by using the $i$QIST software package~\cite{iqist,iqist_new}, which implements the hybridization expansion CT-QMC algorithm~\cite{Gull}. The calculations are done on inverse temperature $\beta = 1000.0$, and the Legendre orthogonal polynomial representation is adopted to obtain high-precision GFs~\cite{lewin}. The excellent agreement shows that the NO-based Lanczos method provides an accurate and efficient impurity solver that has the potential of wide application in the DMFT studies.
\begin{figure}[t]
\vspace{-2.0cm}
\centering
   \includegraphics[width=0.7\textwidth]{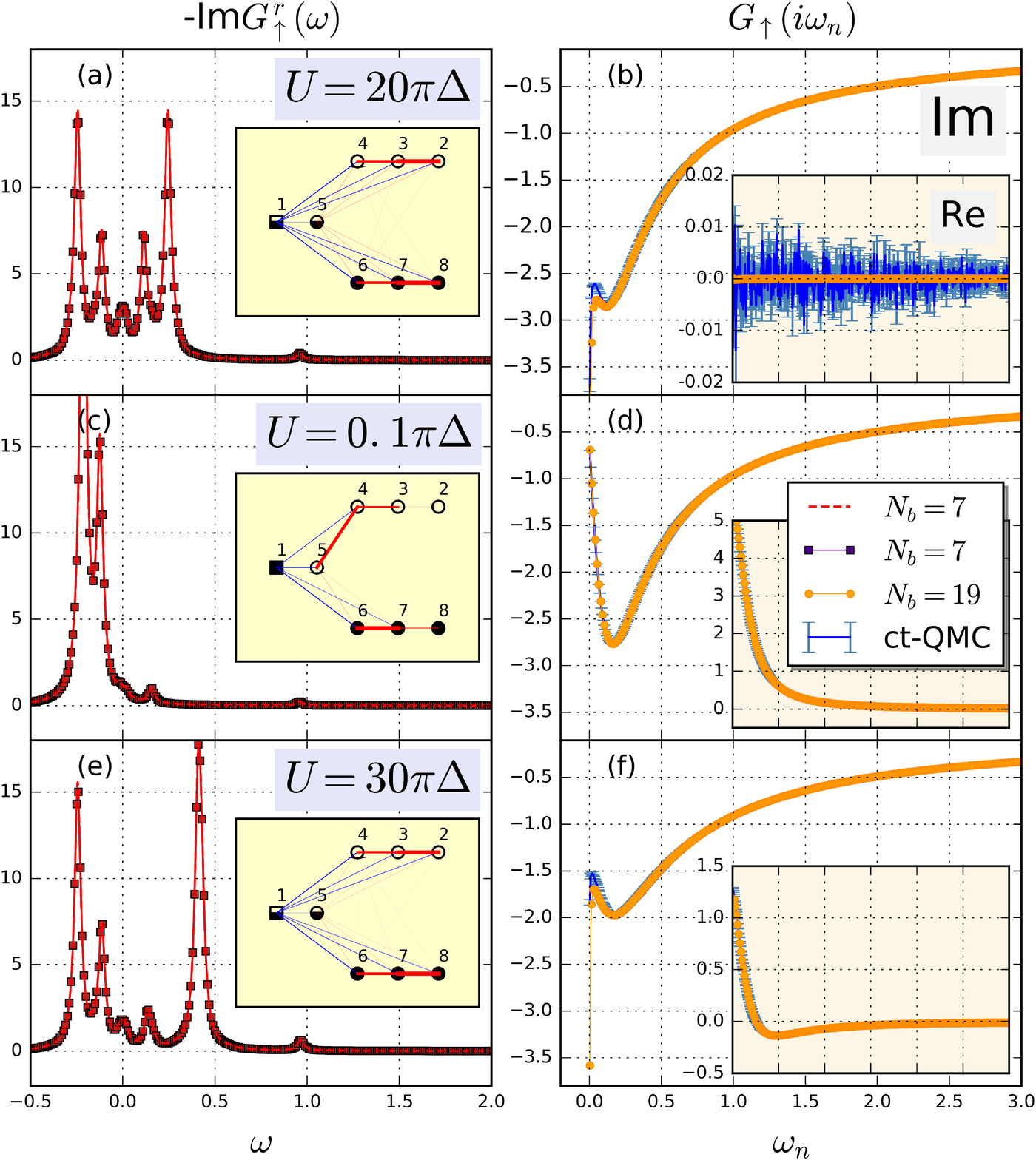}
\caption{(Color online). Zero-temperature retarded GFs $\text{Im}G^{r}_{\uparrow}(\omega)$(left) and the Matsubara GFs $G_{\sigma}(i \omega_n)$(right) of single-impurity AIM for various parameters. (a) and (b): $U=2 \mu$, (c) and (d): $U=0.01 \mu$, and (e) and (f): $U=3\mu$. $\mu = 10 \pi \Delta$, $\pi \Delta=0.02$, and $\eta = 0.02$. In the left column, $\text{Im}G^{r}_{\uparrow}(\omega)$'s from NO-based Lanczos (solid squares) are compared with the exact Lanczos results (red dashed lines) for $N_{b}=7$. The insets show the converged NOs and the structure of $H^{(i_0)}$, with the line width proportional to the hopping strength. In the right column, $\text{Im}G_{\uparrow}(i\omega_n)$'s from NO-based Lanczos (solid cycles) for $N_{b}=19$ are compared with the CT-QMC results (blue solid lines with error bars) for continuous bath. The insets give the corresponding $\text{Re}G_{\uparrow}(i\omega_n)$. \label{Fig_MG1}}
\end{figure}
\begin{figure}[t]
\centering
   \includegraphics[width=0.6\textwidth]{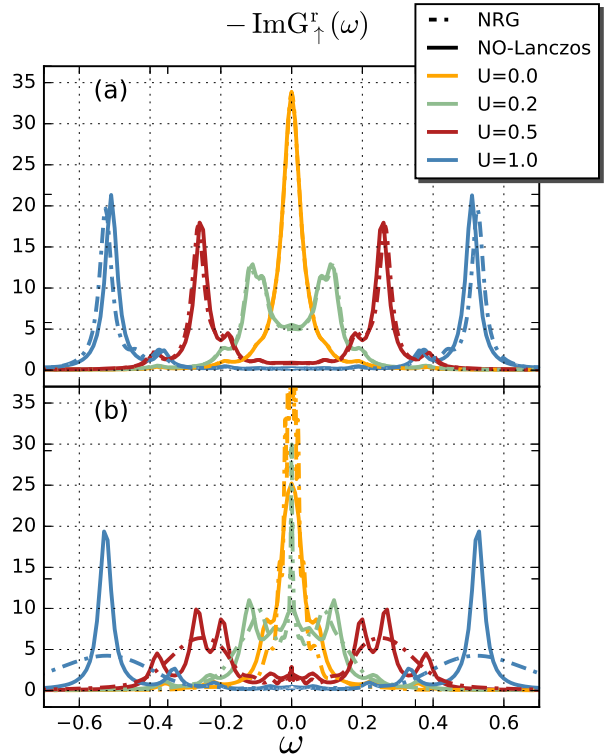}
\caption{(Color online). The spectral function of single-impurity AIM for $U = 0.0$, $0.2$, $0.5$, and $1.0$. NO-based Lanczos results (solid lines) are compared with the NRG results (dashed lines). (a) Results for the Wilson chain Hamiltonian with $N_{b}=19$ bath sites obtained from the logarithmic discretization. (b) Results for the continuous bath AIM. NO-based Lanczos calculation uses a discrete Hamiltonian with $N_{b}=19$ fitted bath sites, and NRG uses the semi-infinite Wilson chain Hamiltonian. \label{Fig_green_NRG}}
\end{figure}

\begin{figure}[t]
\centering
   \includegraphics[width=0.6\textwidth]{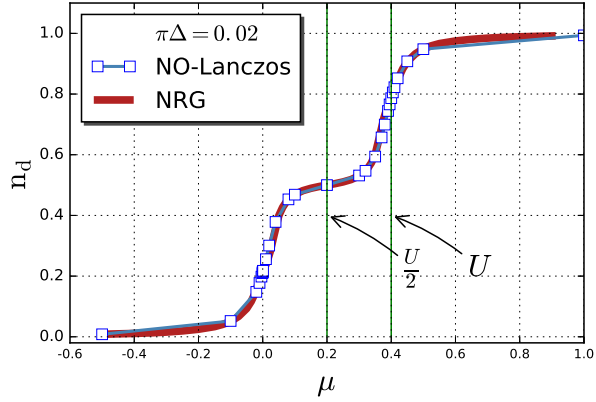}
\caption{(Color online). The impurity occupation $n_d$ as a function of chemical potential $\mu$. The parameters are  $U=0.4$ and $\pi \Delta = 0.02$. \label{Fig_muN} }
\end{figure}

\begin{figure}[t]
\centering
   \includegraphics[width=0.5\textwidth]{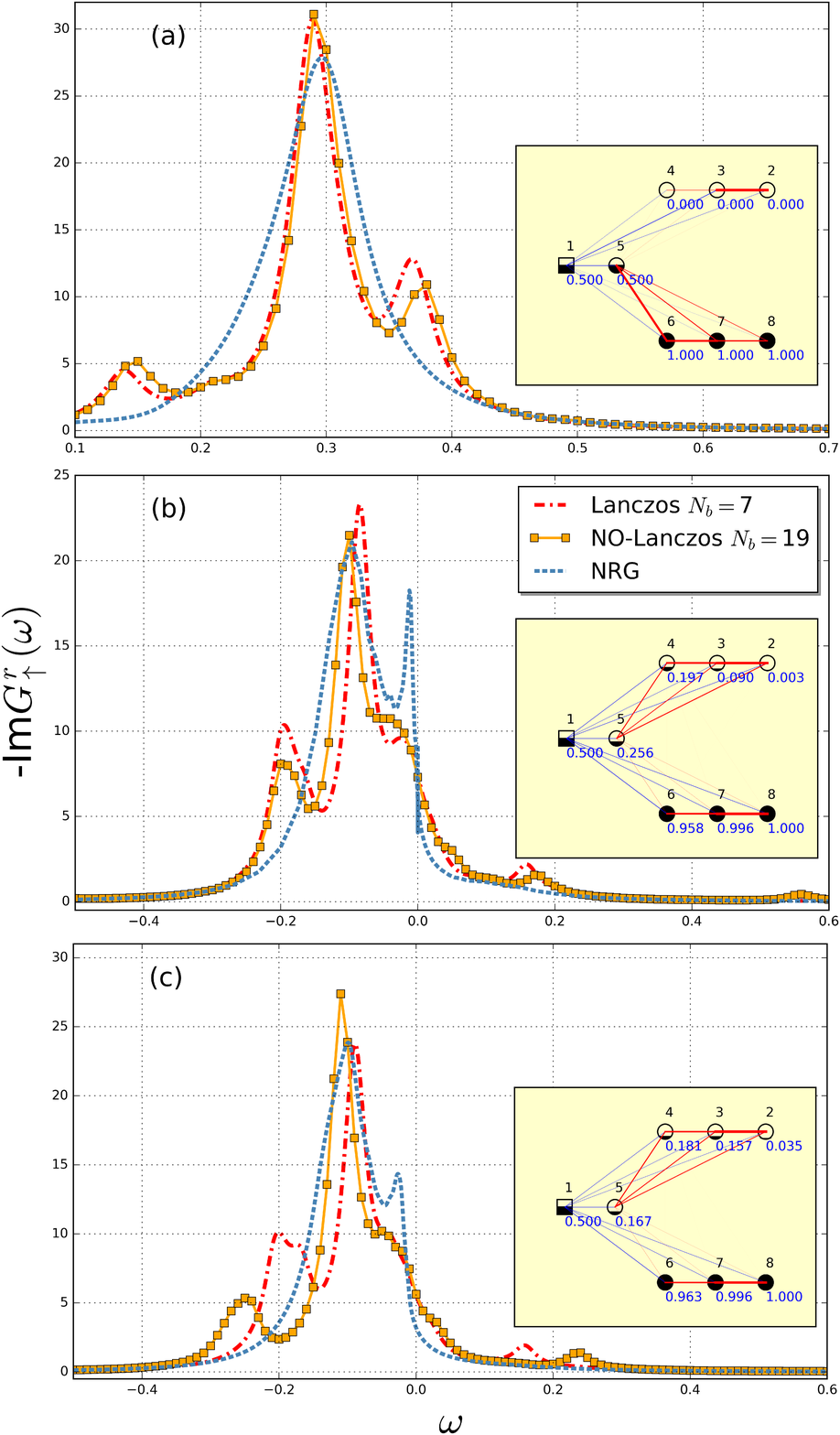}
\caption{(Color online). The spin up GF for single impurity model with local magnetic field. (a) $U=0.4$, $h=0.1$; (b) $U=0.2$, $h=-0.006$; and (c) $U=0.2$, $h=-0.0122$. Here $\mu=U/2$, and $\pi \Delta=0.02$.
The inset shows the structure of the converged Hamiltonian, with the line width proportional to the hopping strength. \label{Fig_G_magnetic}}
\end{figure}

\begin{figure}[t]
\vspace{-3.0cm}
\centering
   \includegraphics[width=0.6\textwidth]{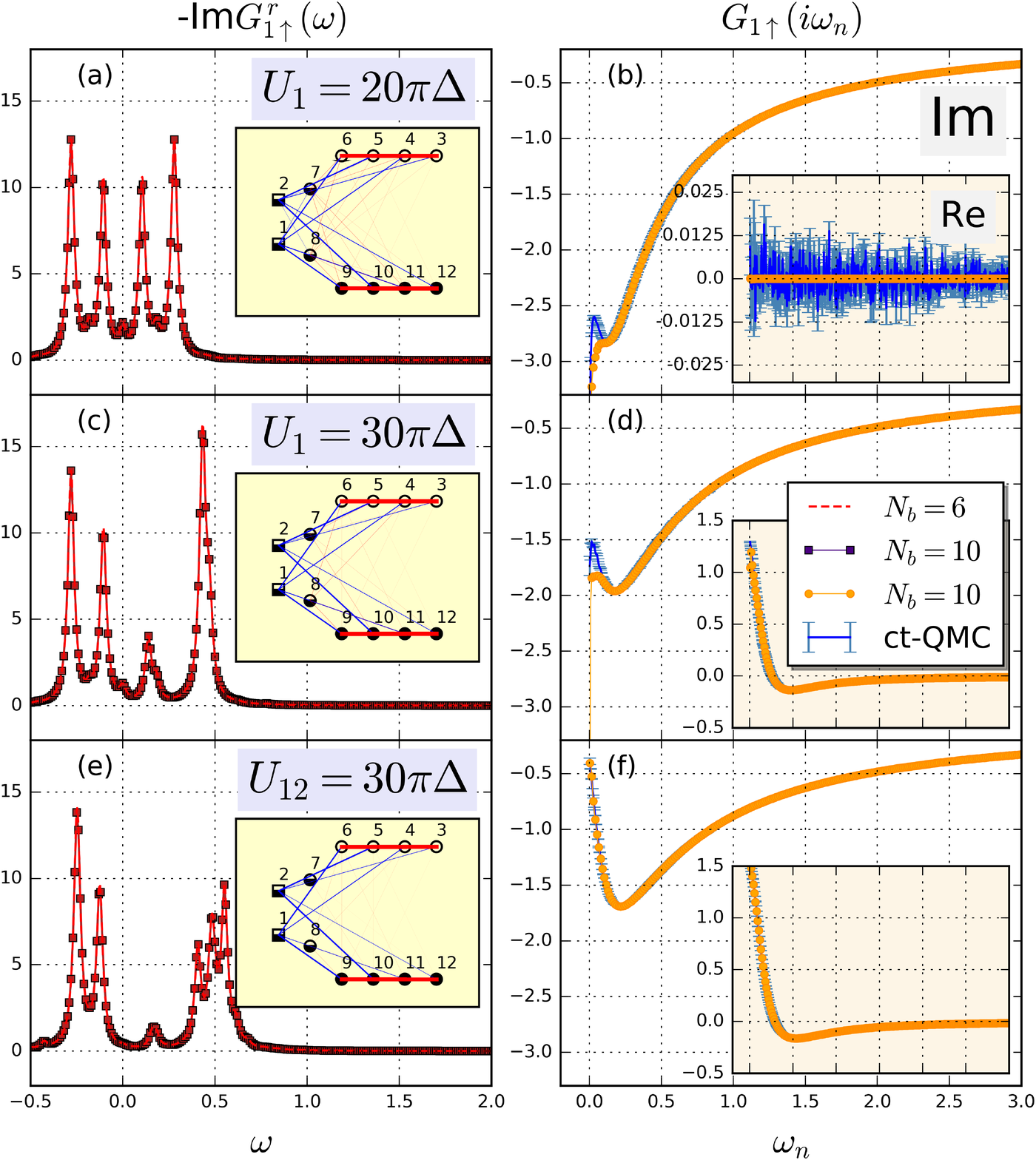}
\caption{(Color online). Zero-temperature retarded GFs $\text{Im}G^{r}_{1\uparrow}(\omega)$ (left) and the Matsubara GFs $G_{1\sigma}(i \omega_n)$ (right) of the two-impurity AIM with diagonal hybridization for various parameters. (a) and (b): $\mu_1 = 10 \pi \Delta, U_1 = 2 \mu_1, U_{12}=0$; (c) and (d): $\mu_1 = 10 \pi \Delta, U_1 = 30 \pi \Delta < 2 \mu_1, U_{12}=0$; and (e) and (f): $\mu_1 = 10 \pi \Delta, U_1 = 2 \mu_1, U_{12}=30 \pi \Delta$. Other parameters are $\mu_2 = 20 \pi \Delta$,$U_2 = 2 \mu_2$, and $\pi \Delta=0.02$. The broadening parameter is $\eta=0.02$. In the left column, $\text{Im}G^{r}_{1\uparrow}(\omega)$'s from NO-based Lanczos (solid squares) are compared with the exact Lanczos results (red dashed lines) for $N_{b}=6$. The insets show the converged NOs and the structure of $H^{(i_0)}$, with the line width proportional to the hopping strength. In the right column, $\text{Im}G_{1\uparrow}(i\omega_n)$'s from NO-based Lanczos (solid cycles) for $N_{b}=10$ are compared with the CT-QMC results (blue solid lines with error bars) for continuous bath. The insets give the corresponding $\text{Re}G_{1\uparrow}(i\omega_n)$. \label{Fig_MG2}}
\end{figure}

\begin{figure}[t]
\vspace{-2.0cm}
\centering
   \includegraphics[width=0.5\textwidth]{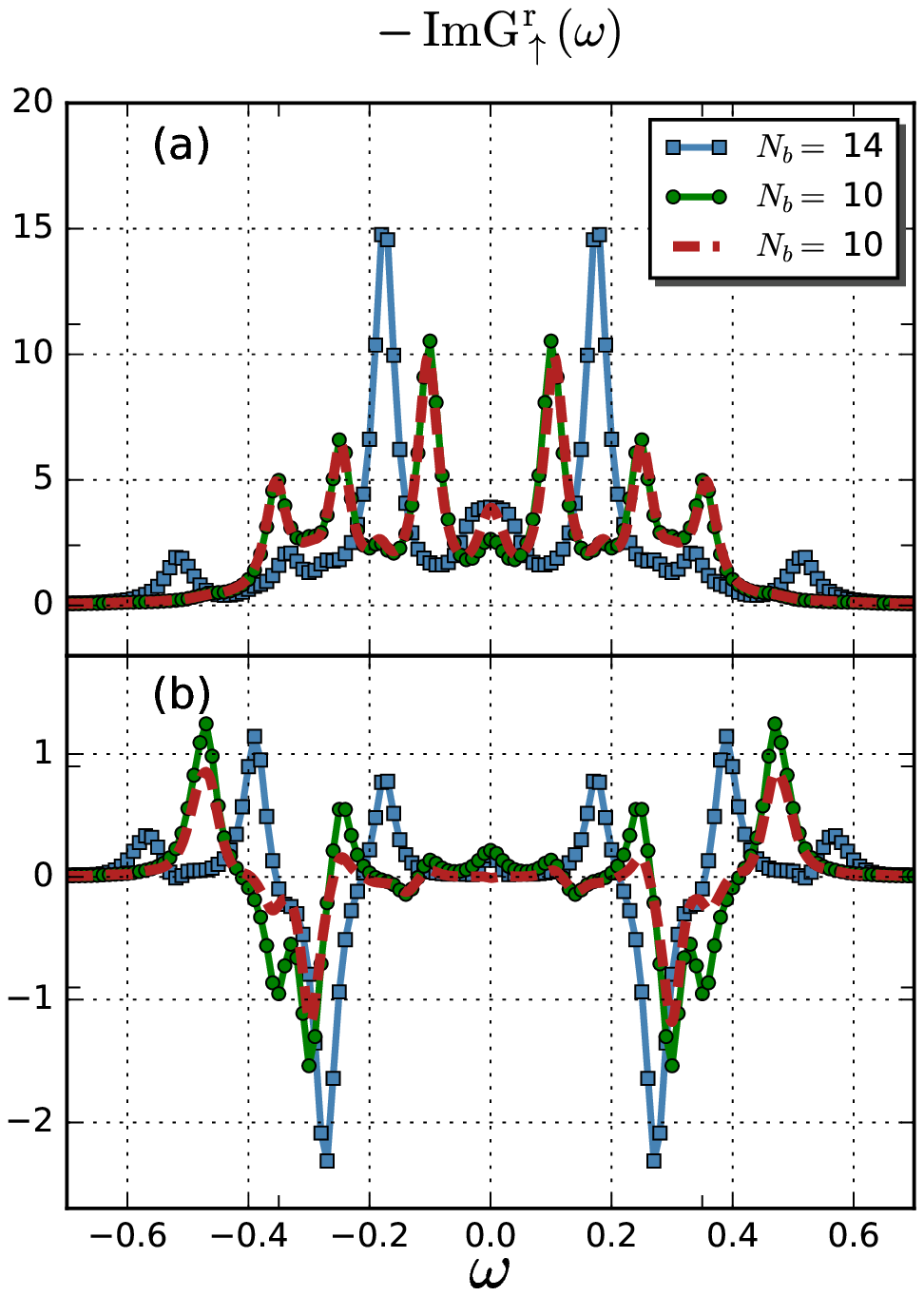}
\caption{(Color online). Zero-temperature retarded GFs  of the two-impurity AIM with both diagonal and off-diagonal hybridizations, (a) diagonal GF $- {\text Im}G_{11 \uparrow}^{r}(\omega)$ and (b) off-diagonal GF $- {\text Im}G_{12 \uparrow}^{r}(\omega)$. Solid squares and dots with guiding lines are from NO-based Lanczos for $N_b=14$ and $N_b=10$, respectively. The dashed lines are from the full Lanczos calculation for $N_b=10$ for comparison. 
Parameters are $\mu_1 = 10 \pi \Delta_1$, $U_1 = 2 \mu_1$, $\mu_2 = 20 \pi \Delta_2$, $U_2 = 2 \mu_2$, $U_{12}=0$, $\pi \Delta_1=0.02$, and $\pi \Delta_2= 0.01$. }  \label{Fig14}
\end{figure}

\subsection{Single-impurity AIM}

\subsubsection{Particle-hole symmetric case}

For the particle-hole symmetric parameter $\mu=U/2$, the results of spectral function and Matsubara GF are shown in Fig.~\ref{Fig_MG1}(a) and (b), respectively. The symmetry is fulfilled in these results, {\it i.e.}, $\text{Im}G^{r}_{\sigma}(\omega) = \text{Im}G^{r}_{\sigma}(-\omega)$ and $\text{Re}G_{\sigma}(i\omega_n) = 0$. Especially the NO and the structure of $H^{(i_0)}$ also bear such symmetry, as shown in the inset of Fig.~\ref{Fig_MG1}(a). In this case, both the $N_{b}=7$ results in Fig.~\ref{Fig_MG1}(a) and the $N_{b}=19$ results in Fig.~\ref{Fig_MG1}(b) agree well with the exact ones from full Lanczos and CT-QMC, respectively. Note that in the small frequency limit, the small deviation in Im$G_{\uparrow}(i \omega_n)$ (main figure of Fig.~\ref{Fig_MG1}(b) and (f)) from the CT-QMC data is probably due to the error in fitting the hybridization function with $N_b=19$ bath sites in our NO-Lanczos calculation.

We also made comparison between NO-based Lanczos and NRG which is supposed to be one of the most accurate method for AIM. In Fig.~\ref{Fig_green_NRG}(a), we compare the spectral functions from NO-based Lanczos and NRG for the same Wilson chain Hamiltonian of AIM with $N_b=19$ bath sites obtained from the logarithmic discretization. In NRG, a truncation of the high energy eigenstates are carried out to avoid the exponential increase of the Hilbert space. The good agreement from small to large $U$ shows that NO-based Lanczos method accurately produces the spectral function of the AIM with $N_{b}=19$. 

In Fig.~\ref{Fig_green_NRG}(b), we compare the spectral function of the AIM with a continuous bath obtained from the NO-based Lanczos and that from NRG. For the former, we use $N_{b}=19$ bath sites and bath parameters are fitted from the Lorentzian hybridization function. The NRG results are for a semi-infinite Wilson chain. Here, the apparent significant difference comes from the discretization error of representing a continuous bath by $N_{b}=19$ discrete bath sites. Qualitative agreement in the features such as the Kondo resonance and the upper/lower Hubbard peaks is observed. Note that the high energy incoherent Hubbard peaks are over broadened in NRG. This shows that $N_{b}=19$ is not sufficient for a quantitative calculation of the spectral density. 

\subsubsection{Particle-hole asymmetric case}

For AIM at particle-hole asymmetric point $\mu \neq U/2$, we made comparisons for weak interaction $U=0.01\mu$ and strong interaction $U=3.0\mu$, respectively in Fig.~\ref{Fig_MG1}(c)-(d), and Fig.~\ref{Fig_MG1}(e)-(f). The agreement in the spectral function with the exact Lanczos for a small cluster $N_{b}=7$, and in the Matsubara GF for $N_{b}=19$ with the CT-QMC results are very good. The only notable discrepancy appears in the small frequency regime in $G_{\uparrow}(i \omega_n)$ where the error bar of CT-QMC data is relatively large.
Based on the comparison, we confirm that the NO-based Lanczos algorithm is also applicable to the particle-hole asymmetry case only by changing the initial space in state \ref{initial space}) of the sparse Lanczos process. Specifically, we include all the configurations of cluster and same configurations of valence and conduction baths as in the symmetric case.
 
Taking the single impurity model as an example. For the particle-hole symmetric case, the initial space $S^{0}$ contains only two Slater determinants $|\phi_1 \rangle^{0}$ and $|\phi_2 \rangle^{0}$ of Fig.~\ref{Fig_extend_iteration}. They represent the states with one electron on the impurity or on the other site of the cluster, with fully occupied valence and empty conduction orbitals.
For the asymmetric case, $S^{0}$ should contain all $4^{2}=16$ different configurations of the two cluster sites, with fully occupied valence and empty conduction orbitals.

To obtain a global view on the performance of NO-based Lanczos away from the particle-hole symmetry, we plot the $n_d - \mu$ curve for the AIM in Fig.~\ref{Fig_muN} and compare it with that obtained from NRG. For a fixed hybridization strength $\pi \Delta = 0.02$, a quantitative agreement with NRG is obtained in the whole range of $\mu$. This shows that the NO-based Lanczos is applicable also the particle-hole asymmetric case.
 
\subsubsection{Under magnetic field}

We also study the influence of a local magnetic field on the local spectral function using the NO-Lanczos method, described by the following Hamiltonian,
\begin{equation}
\hat{H} = \hat{H}_{cond} + \hat{H}_{imp} +\hat{H}_{hyb} + 2h \hat{S}_z,
\end{equation}
where $	\hat{S}_z = ( n_{d \uparrow} - n_{d \downarrow} ) /2$ is the impurity spin-$z$ operator. As the spin up and down density matrix is treated separately in the NO-Lanczos, the algorithm is naturally applicable for this case. The GF is shown in Fig.~\ref{Fig_G_magnetic} for three different parameters. Compared to NRG results for a continuous bath, reasonable agreement is obtained for the NO-Lanczos results with $N_{b}=19$. For the smaller $U$ cases in Fig.~\ref{Fig_G_magnetic} (b) and (c), Kondo peak appears near the Fermi energy but shifted to $\omega = h$. We can see the overall agreement in the peak position and line shape. In the NO-Lanczos results, the Kondo peak is not as sharp as those in NRG, presumably due to insufficient number of bath sites near the Fermi energy.

\subsection{Two-impurity AIM}

For the two-impurity AIM, we first use the diagonal matrix form of the hybridization function of Eq.~(\ref{hybrid Lorentzian}). The NO-based Lanczos algorithm for the two-impurity AIM is the same as that for the single-impurity AIM. In the calculation, it is found that there are two partially occupied bath sites instead of one as in the single-impurity case. We first checked our code with the case $U_{12}=0$, where the two-impurity AIM is reduced to two decoupled single-impurity AIMs. In this case, the NO-Lanczos method correctly produces GFs identical to those of the single-impurity AIM, as shown in Fig.~\ref{Fig_MG2}(a) and (b) for the particle-hole symmetric case, and in Fig.~\ref{Fig_MG2}(c) and (d) for the asymmetric case. Applying the NO-based Lanczos method to the non-trivial $U_{12} \neq 0$ case and comparing the results with conventional Lanczos for $N_{b}=6$ (Fig.~\ref{Fig_MG2}(e)), and with CT-QMC for $N_b=10$ (Fig.~\ref{Fig_MG2}(f)), we again obtain excellent agreement. 
In the main figure of Fig.~\ref{Fig_MG2}(b) and (d), Im$G_{1 \uparrow}(i \omega_n)$ has small deviations from the CT-QMC data in the small Matsubara frequency regime. They are due to errors in fitting the hybridization function with $N_b=10$ bath sites in our NO-Lanczos calculation. In the insets of Fig.~\ref{Fig_MG2}(b), (d), and (f), Re$G_{1 \uparrow}(i \omega_n)$ agrees well with the CT-QMC data. 

For the two-impurity AIM with both diagonal and hybridization, we uses Eq.(\ref{hybrid matrix}) with 
\begin{eqnarray} \label{off-diagonal}
 && \Gamma_{11}(i\omega_n)=  \Gamma_{22}(i\omega_n) = \frac{\pi \Delta_{1} \omega_c}{i \omega_n + i \omega_c \text{sgn}(\omega_n)},  \nonumber \\
&& \Gamma_{12}(i\omega_n)= \Gamma_{21}(i\omega_n)=  \frac{\pi \Delta_{2} \omega_c}{i \omega_n + i \omega_c \text{sgn}(\omega_n)}.
\end{eqnarray}
Here $\Delta_1$ and $\Delta_2$ control the diagonal and the off-diagonal hybridization strength, respectively. For $\Delta_2 \neq 0$, a negative sign problem occurs in the CT-QMC simulation and hampers the production of reliable results. In Fig.~\ref{Fig14}, we therefore compare the NO-based Lanczos results with the full Lanczos result for $N_b=10$. The parameters of AIM are fitted from Eq.(\ref{off-diagonal}) by assigning $3$ bath sites to each of the two diagonal baths and $4$ to the off-diagonal bath, respectively. Good agreement is obtained for the diagonal GF. For the off-diagonal GF, the correct peak positions are produced but the the height of certain peaks are less accurate. In Fig.~\ref{Fig14}, we also show the NO-based Lanczos results for $N_b=14$.

\section{Conclusion\label{Conclusion}}  
In this paper, we studied the NO-based Lanczos algorithm for calculating the ground state and zero-temperature GFs of AIMs, following the algorithm proposed by Lu~\cite{Lu}. We provide technical details and performance analysis of this algorithm that are important but lacking in the original literature, and confirmed a key feature of this algorithm, {\it i.e.} the number of partially occupied bath sites is equal to the number of impurity sites. We observe that the computational complexity is proportional to the number of bath sites up to $N_{b}=27$ using error $\epsilon_f = 3 \times 10^{-7}$, although this dependence is expected to become $N_b^{2}$ in the large $N_b$ limit. It is noted that in Ref.~\cite{He2}, the complexity is proportional to $N_b^3$ for a different algorithm. We also extend the algorithm to the cases of particle-hole asymmetry, under a local magnetic field, and of two impurities with both diagonal and off-diagonal hybridization. Our results are compared to the full Lanczos, NRG, and CT-QMC results, all giving excellent agreement. Our results show that the weak-entanglement feature of the ground state of AIMs can be employed successfully to reduce the computational complexity and renders AIMs with $N_b \sim 10^{2}$ to be solved accurately within $\mathcal{O}(10^{4})$ Slater determinants, therefore demonstrating that NO-based Lanczos is a promising impurity solver for wide applications in DMFT. At present, due to technical reasons in our coding process, we could only process up to $N_b = 30$ bath. This is because integer is used to index the orbital in our program, and the maximum of integer in Fortran is $4^{32}$. Further extension of our code to study AIMs with $N_{b} \geqslant 30$ and $N_{d} \sim 5$ is straightforward and in progress.

\section*{Acknowledgement}
This work is supported by 973 Program of China (2012CB921704), NSFC grant (11374362), Fundamental Research Funds for the Central Universities, and the Research Funds of Renmin University of China 15XNLQ03. LH was supported by the Natural Science Foundation of China (No.~11504340), the Foundation of President of China Academy of Engineering Physics (No.~YZ2015012), and the Science Challenge Project of China (No.~TZ2016004).

\end{document}